\documentclass[aps,pre,reprint,longbibliography,superscriptaddress,groupaddress,floatfix]{revtex4-1}
\usepackage{graphicx}
\usepackage{dcolumn}
\usepackage{bm}
\usepackage{amsmath}
\usepackage{amsfonts}
\usepackage{amssymb}
\usepackage{graphicx}
\usepackage{xcolor}

\makeatletter
\DeclareFontFamily{OMX}{MnSymbolE}{}
\DeclareSymbolFont{MnLargeSymbols}{OMX}{MnSymbolE}{m}{n}
\SetSymbolFont{MnLargeSymbols}{bold}{OMX}{MnSymbolE}{b}{n}
\DeclareFontShape{OMX}{MnSymbolE}{m}{n}{
    <-6>  MnSymbolE5
   <6-7>  MnSymbolE6
   <7-8>  MnSymbolE7
   <8-9>  MnSymbolE8
   <9-10> MnSymbolE9
  <10-12> MnSymbolE10
  <12->   MnSymbolE12
}{}
\DeclareFontShape{OMX}{MnSymbolE}{b}{n}{
    <-6>  MnSymbolE-Bold5
   <6-7>  MnSymbolE-Bold6
   <7-8>  MnSymbolE-Bold7
   <8-9>  MnSymbolE-Bold8
   <9-10> MnSymbolE-Bold9
  <10-12> MnSymbolE-Bold10
  <12->   MnSymbolE-Bold12
}{}

\let\llangle\@undefined
\let\rrangle\@undefined
\DeclareMathDelimiter{\llangle}{\mathopen}
{MnLargeSymbols}{'164}{MnLargeSymbols}{'164}
\DeclareMathDelimiter{\rrangle}{\mathclose} {MnLargeSymbols}{'171}{MnLargeSymbols}{'171}
\makeatother

\begin{document}

\title{Nonequilibrium statistics of harmonically trapped run-and-tumble particles: An exact convolution approach}

\author{Francisco J. Sevilla}
\email[]{fjsevilla@fisica.unam.mx}
\thanks{Corresponding author}
\affiliation{Instituto de Física, Universidad Nacional Autónoma de México,
Ciudad de México, C.P. 04510, México.}

\author{Jair A. \surname{Meléndez Mora}}
\affiliation{Instituto de Física, Universidad Nacional Autónoma de México,
Ciudad de México, C.P. 04510, México.}

\date{Today}

\begin{abstract}
We study one-dimensional run-and-tumble particles confined by a harmonic potential and coupled to an equilibrium thermal bath. Exploiting the coupling of active and thermal degrees of freedom through the Ornstein-Uhlenbeck propagator, we show that the position distribution factorizes, as a convolution, into the Ornstein-Uhlenbeck distribution and the distribution of the athermal run-and-tumble problem. This yields closed-form results, correcting expressions from earlier treatments by identifying the Ornstein-Uhlenbeck propagator, rather than the free-diffusion one, as the correct kernel. Results are confirmed both in Fourier space and by direct Langevin simulation. We further focus our analysis on the stationary regime, characterized by two dimensionless parameters—the ratio of trapping to persistence length, and the ratio of thermal to active diffusion—which control the crossover from a non-Gaussian, boundary-peaked distribution to the Gaussian, equilibrium-like limit. Energy fluctuations and the Kullback-Leibler divergence from the equilibrium distribution quantify the resulting non-equilibrium character of the confined active particle.
\end{abstract}


\maketitle

\section{Introduction}
The run-and-tumble model (RTM) of active motion \cite{SchnitzerPhysRevRE1993}, belongs to the class of mathematically simple models that describes the ubiquity of non-Brownian transport. The model explicitly describes the stochastic dynamics among the particle's self-propelling states that define the characteristics of active motion. In the one-dimensional case, the intermittent change between the two possible directions of motion (right and left), at random time intervals, describes the particle's motion. The model is equivalent to the Poisson-Kac process and to the telegrapher equation, previously considered to describe the finite speed propagation of the particle-position probability density. In the context of active motion, this model describes \emph{persistent motion}, i.e., the tendency to preserve the direction of motion, a feature that exhibits the nature of nonequilibrium under the effects of confinement. In the simple case where no bias in the direction of motion is present, the transitions rates between the self-propulsion states are equal and constant $\alpha/2$. In addition we also assume that the running speed does not depends on the direction and is constant $v$.

Under conditions of confinement, it has proven that active motion reveals its underlying non-equilibrium dynamics that characterize it. These effects have been of wide interest, particularly for the motion of 1-dimensional run-and-tumble particles (RTPs) restricted by the harmonic potential  \cite{SevillaPRE2019,DharPhysRevE2019,Garcia-MillanJStatMech2021,CrisantiPhysRevE2023,MeyerNewJPhys2023,FrydelPoF2023,DuttaPhysRevE2024,GueneauPhysRevE2026}. Certainly, the interplay between activity and spatial constraints clearly exposes the intrinsic non-equilibrium character of active matter \cite{SmithPhysRevE2023,FaragoPhysRevE2024,SinghPhysRevE2025}. The analysis of 2-dimensional active motion restricted by the harmonic potential has also been of wide interest and both, experimental and theoretical studies, have recently appeared  \cite{TakatoriNatcomms2016,ArgunPRE2016,WexlerPhysRevResearch2020,SchmidtNatureComm2021,ButtinoniEPL2022,CaraglioPhysRevLett2022,SmithPhysRevE2022,SzamelPhysRevE2023,BaldovinPhysRevLett2023,FrydelPoF2024,SunPhysRevE2025,GueneauPhysRevE2026}. 

In this paper, we investigate one-dimensional independent run-and-tumble particles subjected to a harmonic restoring force toward the origin and thermal fluctuations arising from an equilibrium environment at temperature $T$. The confining potential induces a nontrivial coupling between the fluctuations due to the active motion and those due to thermal noise, making this otherwise minimal model, a fertile testing ground for nonequilibrium statistical mechanics and for the assessment of analytical approaches beyond equilibrium paradigms \cite{Garcia-MillanJStatMech2021}. Thus, building on standard statistical mechanics foundations, we introduce a conceptually transparent and technically direct route to exact solutions. This framework yields closed-form results without resorting to auxiliary constructions or advanced formalisms that have previously been invoked to bypass the analytical difficulties of the problem, thereby providing genuinely new insights into the structure of confined active matter systems.

Briefly, in this paper we show that two dimensionless parameters define the non-Gaussian steady state of the particle positions distribution, namely, $\ell$ the ratio of the persistence length to the trapping length, and $\mathcal{D}$, the ratio of the thermal diffusion coefficient to the effective diffusion coefficient of active motion that emerges from the randomization of the direction of motion. The energy fluctuations and the Kullback-Leibler divergence between the exact stationary distribution of the particle positions and the corresponding one of equilibrium, is analyzed.

The paper is organized as follow. In section \ref{sect:Model} we present the theoretical framework of our study. In section \ref{sect:Solution} we focus in the system stationary solutions and their implications for the system energy fluctuations and for the information entropy. We finally give our concluding remarks in section \ref{sect:Conclusions}

\section{\label{sect:Model} Run-and-tumble motion confined by a harmonic trap and subject to thermal fluctuations}
In this section we consider run-and-tumble particles moving in one dimension and trapped by the harmonic potential $U(x)=\kappa x^{2}/2$, 
where $\kappa$ is a constant that characterizes the intensity of the trapping. Although the coupling of the active particles with the surrounding medium is in general complex (as there are different mechanisms that may influence the motion as chemotaxis, phototaxis, gravitaxis, etc.), here we analyze the simple situation for which the particle swims in a viscous fluid at low Reynolds numbers, at the constant speed $v>0$ and tumbling at the constant rate $\alpha/2>0$, giving rise to the \emph{persistent length} $l_\text{pers}=v/(\alpha/2)$. In this regime, a description of the motion based on an overdamped dynamics is adequate.

The position of the particle $x(t)$ is a stochastic process whose dynamics is given by the stochastic differential equation 
\begin{equation}\label{Langevin}
\frac{d}{dt}x(t)=-\mu\kappa x(t)+v\sigma(t)+\xi(t),
\end{equation}
which we simply refer to as the ``Langevin equation'' for $x(t)$. A complete description of the process is achieved by making explicit the stochastic processes $\sigma(t)$ and $\xi(t)$, which corresponds to a stationary Markov dicotomic (telegraphic) process that transits between the values $+1$ and $-1$ with the same rate $\alpha/2$, $\xi(t)$ is Gaussian white noise, i.e. $\langle\xi(t)\rangle=0$ amd $\langle\xi(t)\xi(s)\rangle=2D_T\delta(t-s)$ with $D_T$ the diffusion constant. A simple version of the Euler-Maruyama method was implemented to numerically integrate Eq.~\eqref{Langevin}. Thus we can interpret the effects of the trapping potential as modifying the run-and-tumble particle velocity according to  
\begin{subequations}\label{SwimmingSpeeds}
\begin{align}
v_{R}(x)&=v-\mu\, \kappa\, x,\label{RightSwimmingSpeed}\\
v_{L}(x)&=v+\mu\, \kappa \, x,\label{LeftSwimmingSpeed}
\end{align}
\end{subequations}
which give the effective non-negative right and left space-dependent swimming speeds, respectively. $\mu>0$ is the mobility of the particle in the fluid, which for simplicity is assumed to be space independent and finite, with physical units of [Time/Mass]. Notice that both $v_{R}(x)$ and $v_{L}(x)$ are bounded and monotonous functions of the particle position, vanishing at the characteristic \emph{trapping length}:  $l_\text{trap}=v/\mu\kappa>0$, which defines ``soft" boundaries at the $+l_\text{trap}$ and $-l_\text{trap}$ of active motion, respectively. Importantly, we also consider the physical situation of particle sizes for which the thermal fluctuations due to the environment are not negligible, these are embodied in the translational diffusion coefficient $D_{T}=\mu k_B T>0$, being $k_B$ the Boltzmann constant and $T$ the fluid's absolute equilibrium temperature.
\begin{figure}
\includegraphics[width=\columnwidth, trim=0 20 0 0]{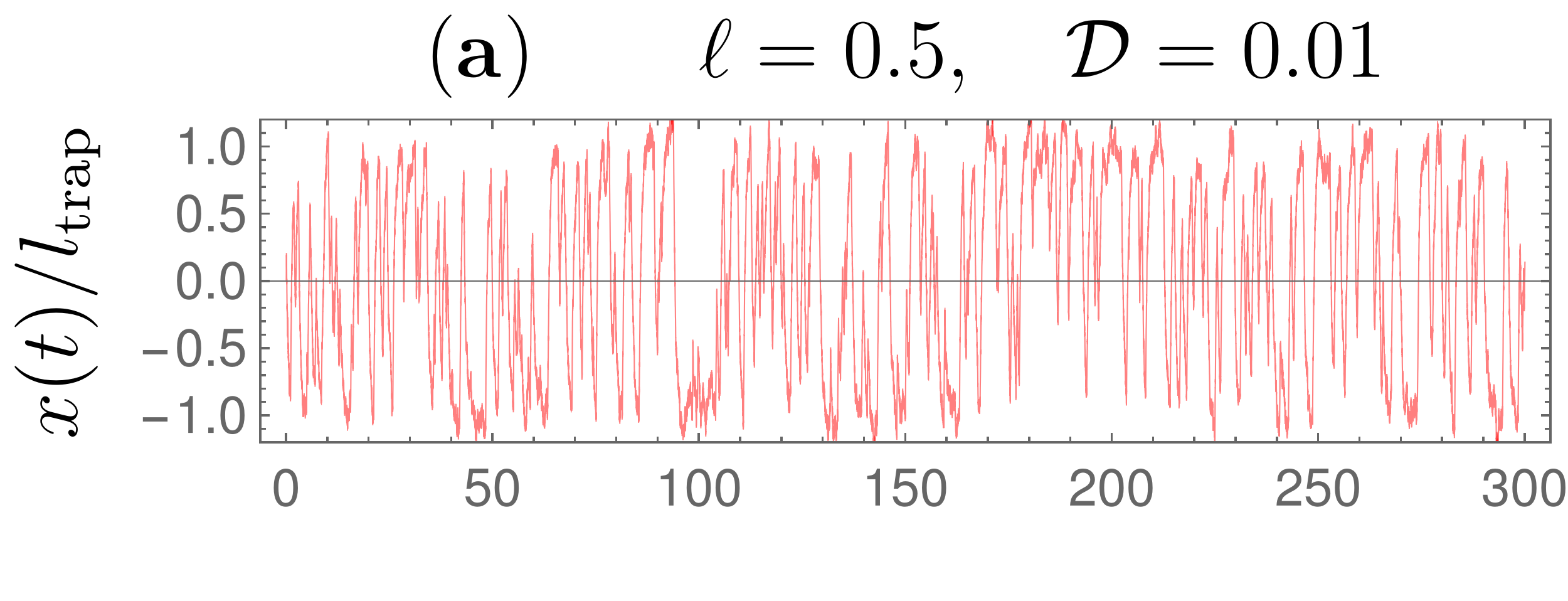}\\
\includegraphics[width=\columnwidth, trim=0 17 0 0]{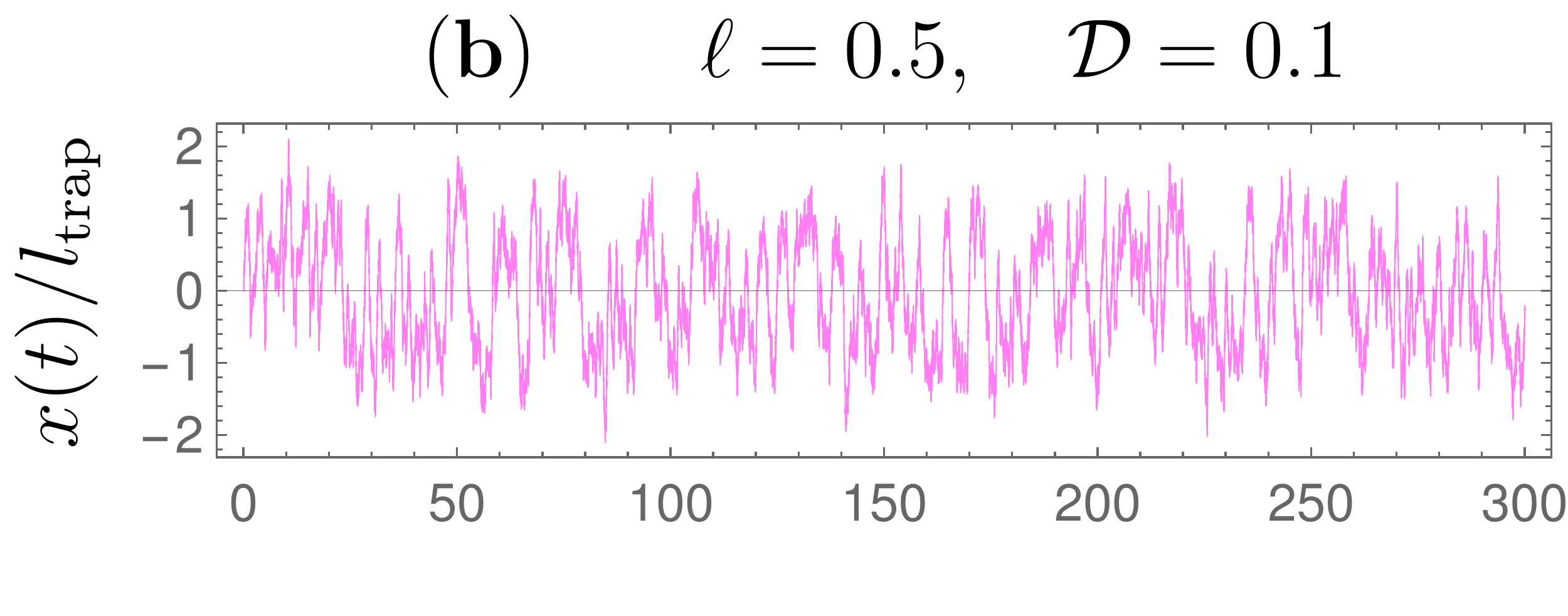}\\
\includegraphics[width=\columnwidth, trim= 0 18 0 0]{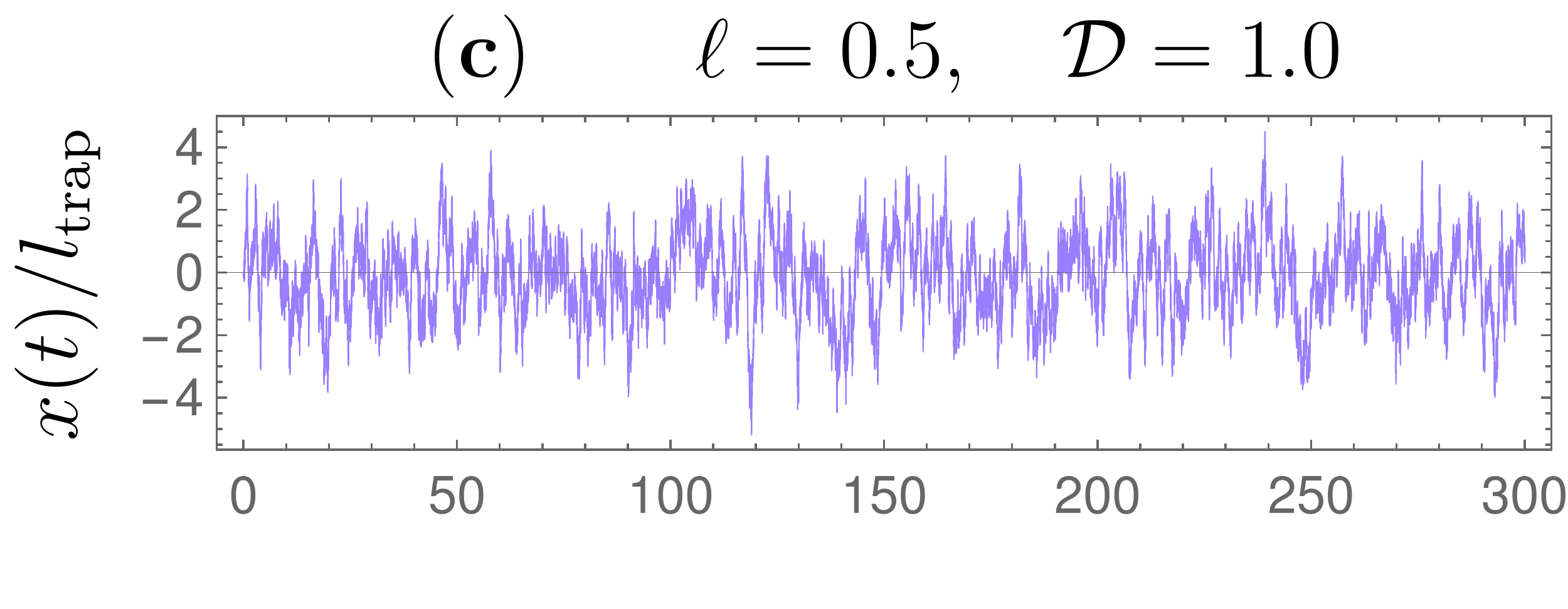}\\
\includegraphics[width=\columnwidth,trim= 0 15 0 0]{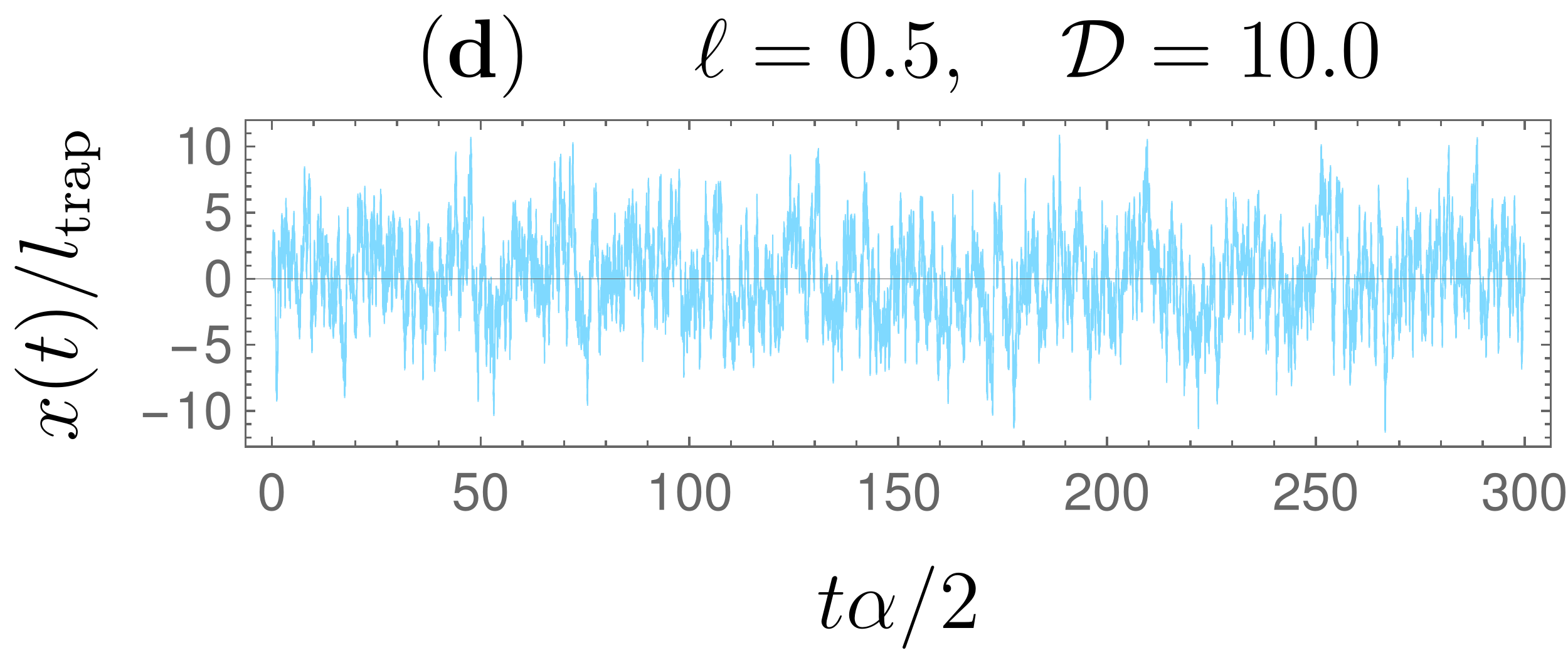}
\caption{Four trajectories of a run-and-tumble particle in the persistent regime are shown for $l_\text{pers}=2l_\text{trap}$($\ell=0.5$) and different values of $\mathcal{D}$. In each case, the dimensionless position of the particle $x(t)/l_\text{trap}$ is shown as function of the dimensionless time $t\alpha/2$. In panel (a), $\mathcal{D}=0.01$, thermal noise is small and the particle dwells around $\pm l_\text{trap}$. As thermal fluctuations increase, the correlations of active motion fade and the particle position distributes around the center of the trap.}
\label{fig:trajectories}
\end{figure}

Two relevant dimensionless parameters rise in our analysis: one given by $\ell\equiv \alpha/2\mu\kappa$, that equivalently gives the ratio $l_\text{trap}/l_\text{pers}$, which measures the competition between the characteristic length of persistence against the characteristic length of trapping. Thus, $\ell\ll1$ means that the effects of persistence go beyond trapping, while $\ell\gg1$ means that persistence is not relevant. The other dimensionless parameter is given by 
$\mathcal{D}\equiv D_T(\mu\kappa/v^2)=D_T/(\mu\kappa l_\text{trap}^2)$, with $v^{2}/\mu\kappa=\mu\kappa l_\text{trap}^2$ is a characteristic diffusion coefficient that involves the self-propelling speed $v$, 
the confinement coefficient $\kappa$ and the particles mobility $\mu$. This gives the effective fluctuations of the particle motion induced by thermal noise, self-propulsion and trapping. $\mathcal{D}\ll1$ means that thermal fluctuations are not important on the run-and-tumble dynamics, while $\mathcal{D}\gg1$ means that standard Brownian motion  under trapping is the dominant dynamics. 

Four sample trajectories are shown in Fig.~\ref{fig:trajectories}, these exemplify the persistent regime ($l=0.5$) for the values: $\mathcal{D}=0.01$ in panel (a) the small-thermal noise regime is illustrated, the particle dwells frequently around $\pm l_\text{trap}$, inducing an accumulation around those positions. As $\mathcal{D}$ is increased, $0.1$ (panel (b)), 1.0 (panel (c)) and 10.0 (panel (d)), this feature is lost as the dynamics is overwhelmingly Brownian, the distribution of the particle positions accumulates around the center of the trap. 

The probability densities of being at $x$ at the instant $t$ and moving to the right, $P_R(x,t)$, and to the left, $P_L(x,t)$, satisfy the well-known equations  
\begin{widetext}
\begin{subequations}\label{RT-Eqs}
\begin{align} 
 \frac{\partial}{\partial t}P_{R}(x,t)+\frac{\partial}{\partial x}\left(v-\mu\,\kappa\, x\right)P_{R}(x,t)&=\frac{\alpha}{2}\bigl[P_{L}(x,t)- P_{R}(x,t)\bigr]+D_{T}\frac{\partial^{2}}{\partial x^{2}}P_{R}(x,t),
\\
 \frac{\partial}{\partial t}P_{L}(x,t)-\frac{\partial}{\partial x}\left(v+\mu\,\kappa\, x\right)P_{L}(x,t)&= 
\frac{\alpha}{2}\bigl[P_{R}(x,t)-P_{L}(x,t)\bigr]+D_{T}\frac{\partial^{2}}{\partial x^{2}}P_{L}(x,t),
 \end{align}
\end{subequations}
\end{widetext}
complemented by the initial distributions $P_R^{(0)}(x)=P_R(x,0)$, $P_L^{(0)}(x)=P_L(x,0)$.   
Exact solutions to these equations are not known (except in the case when the thermal fluctuations are absent \cite{DharPhysRevE2019,GueneauPhysRevE2026}), due to the intertwined effects 
of the spatial constriction of the trapping potential and the thermal fluctuations, however accurate perturbative methods of field theory have been applied \cite{Garcia-MillanJStatMech2021}.  The spatial constriction of the trapping potential makes the effects of the persistence of motion and thermal fluctuations, highly intricate, notwithstanding this, we can make progress based on the methodology used in Ref.~\cite{SevillaPRE2019}.

A simple rearrangement of Eqs.~\eqref{RT-Eqs} shows that the \emph{advective derivative}, $\partial_t \pm v \partial_x$, of the probability densities $P_{R,L}(x,t)$ are driven by the Ornstein-Uhlenbeck generator
\begin{equation}
\hat{\mathcal{L}}_\text{OU}=\frac{\partial}{\partial x}\biggl[\mu\kappa x+D_{T}\frac{\partial}{\partial x}\biggr]
\end{equation}
on the one hand, and by the transitions between the directions of motion at rate $\alpha/2$ on the other. This observation suggests that the right and left probability distributions can be written as
\begin{equation}\label{RL_convoluted}
P_{R,L}(x,t)=\int_{-\infty}^{\infty}dx^{\prime} G_\text{OU}(x-x^{\prime},t\vert 0)\, p_{R,L}(x^{\prime},t)
\end{equation}
where the Ornstein-Uhlenbeck propagator
\begin{multline}\label{OU-G}
G_\text{OU}(x,t\vert x_0)=\left[\frac{\kappa\mu}{2\pi D_{T}(1-e^{-2\kappa\mu\, t})}\right]^{1/2}\times\\
\exp\left\{-\frac{\kappa\mu\, (x-e^{-\mu\kappa t}x_0)^{2}}{2D_{T}(1-e^{-2\kappa\mu\, t})}\right\}
\end{multline}
satisfies the Smoluchoswki equation 
\begin{equation}\label{eq:O-U}
\frac{\partial}{\partial t}G_\text{OU}(x,t\vert x_0)=\hat{\mathcal{L}}_\text{OU}G_\text{OU}(x,t\vert x_0),
\end{equation}
with initial condition $G_\text{OU}(x,0|x_0)=\delta(x-x_0)$
and the probability densities $p_{R,L}(x,t)$ satisfy the run-an-tumble master equations
\begin{multline}\label{RT-pEqs}
 \frac{\partial}{\partial t}p_{R,L}(x,t)\pm\frac{\partial}{\partial x}\left(v\mp\mu\,\kappa\, x\right)p_{R,L}(x,t)= \\
\frac{\alpha}{2}\left[p_{L,R}(x,t)-p_{R,L}(x,t)\right],
\end{multline}
with the initial conditions $p_{R,L}(x,0)=P_{R,L}^{(0)}(x)$.

The analysis of the particle positions distribution is formulated in terms of the \emph{coarse-grained} probability density
\begin{equation}\label{CoarseGrained-P}
 P(x,t)=P_{R}(x,t)+P_{L}(x,t),
\end{equation}
which gives the probability density of finding a particle in a position $x$ at a time $t$ independent of the direction of motion $R$ or $L$. This satisfies the \emph{continuity equation} 
\begin{align}\label{Continuity}
 \frac{\partial}{\partial t}P(x,t)+\frac{\partial}{\partial x}J(x,t)=0,
\end{align}
where the total probability current $J(x,t)$ consists of three contributions, namely 
\begin{equation}
 J(x,t)=J_{\text{act}}(x,t)+J_\text{HO}(x,t)+J_T(x,t),
\end{equation}
where $J_{\text{act}}(x,t)\equiv v\bigl(P_R(x,t)-P_L(x,t)\bigr)$ is the contribution to the total probability current induced by the active motion; $J_\text{HO}(x,t)=-\mu\kappa x\, P(x,t)$ the contribution to the probability current generated by the harmonic potential and $J_T(x,t)\equiv-D_T\partial_xP(x,t)$ the contribution to the current due to thermal fluctuations.

Notice that Eqs. \eqref{RT-pEqs} can also be rewritten in the form of the continuity equation for $p(x,t)\equiv p_R(x,t)+p_L(x,t)$ and $j(x,t)$ as
\begin{equation}\label{act-ContinuityEq}
\frac{\partial}{\partial t}p(x,t)+\frac{\partial}{\partial x}j(x,t)=0,
\end{equation}
where now the probability current $j(x,t)$ is given by
\begin{equation}\label{j-definition}
j(x,t)\equiv j_\text{act}(x,t)+j_\text{HO}(x,t),
\end{equation}
with $j_\text{act}(x,t)\equiv v\bigl(p_R(x,t)-p_L(x,t)\bigr)$ and $j_\text{HO}(x,t)\equiv -\mu\kappa x\, p(x,t)$.

\begin{figure*}
\includegraphics[width=1.0\textwidth]{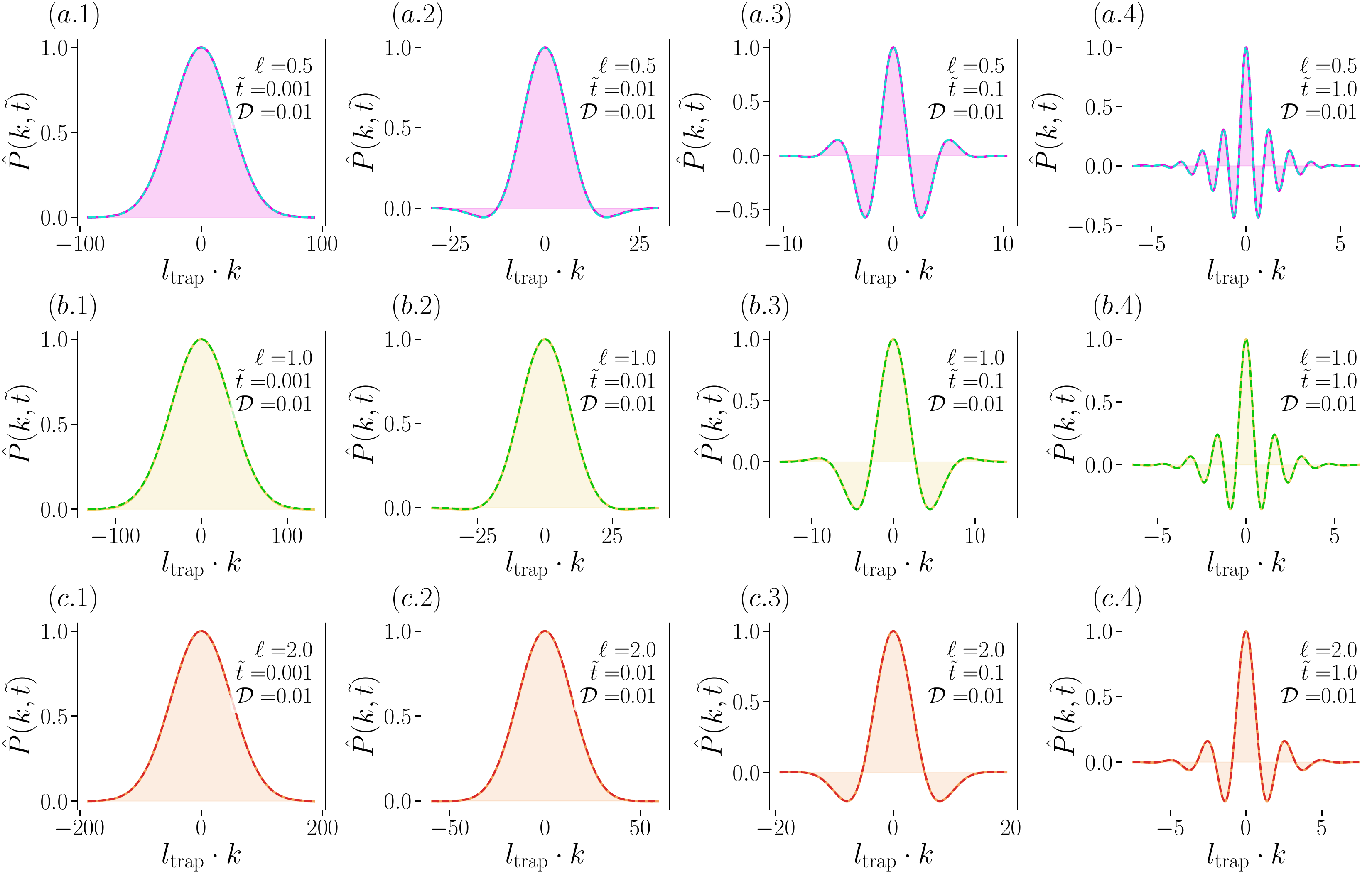}
\caption{\label{fig:FourierConvolutionTime}
Numerical analysis of Eq.~\eqref{Complete-ISF}. $\hat{P}(k,t)$ as function of the dimensionless wavenumber $k\cdot l_\text{trap}$ at different dimensionless times, $\tilde{t}=t \alpha/2$, and for different values of $\ell\equiv l_\text{trap}/l_\text{pers}$, is depicted with solid lines with shaded area underneath.
 This is compared with the product $\hat{G}_\text{OU}(k,t\vert 0)\,\hat{p}(k,t)$ depicted with dashed lines. $\hat{P}(k,t)$ was obtained from Eq.~\eqref{Langevin}, while $\hat{p}(k,t)$ was computed from the athermal dynamics version ($D_T=0$) of Eq.~\eqref{Langevin}. $\hat{G}_\text{OU}(k,t\vert 0)$ was numerically evaluated from Eq.~\eqref{P-OU-Fourier}. Rows $(a)$, $(b)$ and $(c)$ correspond to the values $\ell=0.5$, 1.0, and $2.0$ respectively, while columns 1, 2, 3, 4 correspondingly indicates the values of  $\tilde{t}=10^{-3}$, $10^{-2}$, $10^{-1}$, and $1$. In all panels $\mathcal{D}\equiv D_T/(\mu\kappa\, l_\text{trap}^{2})=0.01$ and wavenumbers are given in units of $l_\text{trap}^{-1}$.}
\end{figure*}
The connection between $J_\text{act}(x,t)$ and $j_\text{act}(x,t)$ is apparent from
Eqs.~\eqref{RT-pEqs}, from which we get
\begin{equation}
 J_\text{act}(x,t)=\int_{-\infty}^{\infty}dx^{\prime}\, G_\text{OU}(x-x^{\prime},t\vert 0)j_\text{act}(x^{\prime},t),
\end{equation}
additionally, $j_\text{act}(x,t)$ satisfies
\begin{multline}\label{ActiveJ}
\frac{\partial}{\partial t}j_\text{act}(x,t)-\frac{\partial}{\partial x}\bigl[\mu\kappa x j_\text{act}(x,t)\bigr]\\
=-\alpha j_\text{act}(x,t)
-v^2 \frac{\partial}{\partial x} p(x,t),
\end{multline}
as can be shown straightforwardly from Eqs.~\eqref{RT-pEqs}. From Eq. \eqref{RL_convoluted} we have that 
\begin{equation}\label{TotalP}
P(x,t)=\int_{-\infty}^{\infty}dx^{\prime} G_\text{OU}(x-x^{\prime},t\vert 0)\, p(x^{\prime},t),
\end{equation}
i.e., such a solution involves the the convolution of the probability density of confined Brownian motion, given by the Ornstein-Uhlenbeck propagator \eqref{OU-G}, and the probability density of confined run-and-tumble motion $p(x,t)$ (see appendix \ref{Proof} for an alternative proof), this is one of our main results which corrects the solution in Ref. \cite{FrydelPoF2023}, where the convolution suggested considers the free-diffusion propagator instead of the correct one, the Ornstein-Uhlenbeck propagator. 

This solution requires the explicit space and time dependence of $p(x,t)$, which to our knowledge, is not known for the general case (the Laplace transform with respect time of $p(x,t)$ has been reported in \cite{DharPhysRevE2019}). However we can proof the validity of Eq.~\eqref{TotalP} in Fourier space. After taking the Fourier transform, $\hat{f}(k)=\int_{-\infty}^{\infty}dx\, e^{-ikx}f(x)$, of \eqref{TotalP} we get the complete \emph{intermediate scattering function} (ISF) 
\begin{equation}\label{Complete-ISF}
\hat{P}(k,t)=\int_{-\infty}^{\infty}dx\, e^{-ikx}P(x,t)=\hat{G}_\text{OU}(k,t\vert 0)\hat{p}(k,t),
\end{equation}
where $\hat{p}(k,t)$ is the ISF of run-and-tumble displacements and
\begin{equation}\label{P-OU-Fourier}
\hat{G}_\text{OU}(k,t\vert 0)=\exp\biggl\{-\frac{D_T}{2\mu\kappa}\bigl(1-e^{-2\mu\kappa t}\bigr)k^2\biggr\}
\end{equation}
is the Fourier transform of the Ornstein-Uhlenbkeck distribution \eqref{OU-G}. $\hat{P}(k,t)$ is the natural
object for characterizing its dynamics because reduces to experimentally
accessible correlators used in dynamic light scattering and differential
dynamic microscopy. 

In Fig.~\ref{fig:FourierConvolutionTime} we present numerical validation of the convolution form of Eq.~\eqref{TotalP} by comparing the left-hand and right-hand sides of \eqref{Complete-ISF} obtained independently by numerical calculations. In each panel the solid line gives  $\hat{P}(k,t)$, computed numerically by generating $10^5$ independent trajectories directly from 
Eq.~\eqref{Langevin}, whereas the dashed line gives the product $\hat{G}_\text{OU}(k,t\vert 0)\,\hat{p}(k,t)$, the second factor was computed numerically from athermal run-and-tumble dynamics 
driven by the same realizations of the telegraphic process $\sigma(t)$ but with $D_T=0$, while $\hat{G}_\text{OU}(k,t\vert 0)$, the
Ornstein-Uhlenbeck kernel \eqref{P-OU-Fourier}, was evaluated numerically.  $\hat{P}(k,t)$ and $\hat{p}(k,t)$ were estimated as normalized histograms on a common grid whose extent was scaled with $\sqrt{\mathcal{D}}$ so as to resolve the distribution for every noise strength. Their Fourier transforms were evaluated through the fast Fourier transform algorithm, zero-padded by a factor of eight to refine the wavenumber grid and multiplied by the phase factor $e^{-ikx_0}$ that accounts for the offset of the histogram domain with respect to the origin. 

The excellent agreement of both curves over the whole range of wavenumbers $k$ is equivalent to the statement given by the solution of Eq.~\eqref{TotalP}. The rows in the figure correspond to increasing values, 0.5, 1.0 and 2.0 of the ratio between the trapping and the persistence lengths, $\ell$, that tunes the system from the strongly persistent, boundary-accumulating regime to the weakly persistent one. The columns correspond to different increasing dimensionless times, $t\, \alpha/2=0.001$, 0.01, 0.1 and 1.0. All curves are evaluated at 
$\mathcal{D}=0.01$ which indicates the persistent regime. 

\section{\label{sect:Solution}The stationary distribution of the particle positions}
The stationary equilibrium-like
solution, $P_\text{st}(x)$, is obtained by setting $J(x,t)=0$ in Eq. \eqref{Continuity}. Equivalently, it can be obtained from \eqref{TotalP} if $p_\text{st}(x)=\lim_{t\rightarrow\infty}p(x,t)$ is known, namely 
\begin{equation}\label{TotalP-Stat}
P_\text{st}(x)=\int_{-\infty}^{\infty}dx^{\prime}G_\text{st}(x-x^{\prime})\, p_\text{st}(x^{\prime}),
\end{equation}
where $G_\text{st}(x)$ denotes the Gaussian distribution 
\begin{equation}\label{Gaussian}
G_\text{st}(x)=\sqrt{\frac{\mu\kappa}{2\pi D_T}}\exp\left\{-\frac{\mu\kappa}{2D_T}x^{2}\right\},
\end{equation}
which corresponds to the sationary distribution of the Ornstein-Uhlenbeck process, a Gaussian centered at the origin and width  $\sigma_\text{ou}=\sqrt{D_T/\mu\kappa}$.

The continuity equation \eqref{act-ContinuityEq} for the active part of motion defines the stationary distribution $p_\text{st}(x)$ by setting $j(x,t)=0$, thus we have from Eq.~\eqref{j-definition} that $j_\text{act}(x)=-j_\text{HO}(x)=\mu\kappa x\, p_\text{st}(x)$. From direct substitution of $j_\text{act}(x)$ in Eq.~\eqref{ActiveJ}, and after some rearrangement we obtain
\begin{equation}\label{EqStatDist}
\frac{\partial}{\partial x}\Bigl[\bigl(v^2-\mu^2\kappa^2 x^2\bigr)p_\text{st}(x)\Bigr]=-\alpha\mu\kappa\, x\, p_\text{st}(x).
\end{equation}
After integration and normalization, we get
\begin{multline}\label{p-act-st}
p_\text{st}(x)=\frac{\Gamma\bigl(\ell+\frac{1}{2}\bigr)}{\sqrt{\pi}\Gamma(\ell)l_\text{trap}}\, \biggl(1-\frac{x^{2}}{l_\text{trap}^{2}}\biggr)^{\ell-1}
\theta\biggl(1-\frac{\vert x\vert}{l_\text{trap}}\biggr),
\end{multline}
where as before, $\ell= \alpha/2\mu\kappa=l_\text{trap}/l_\text{pers}$, and $\theta(x)$ is the Heaviside step function. 
Therefore the complete stationary distribution \eqref{TotalP-Stat} is given explicitly by 
\begin{multline}
\label{StationaryDist}
P_\text{st}(x)=G_\text{st}(x)\frac{\Gamma\bigl(\ell+\frac{1}{2}\bigr)}{\pi\Gamma(\ell)}
\int\limits_{-l_\text{trap}}^{l_\text{trap}}\frac{dy}{l_\text{trap}}
\biggl(1-\frac{y^2}{l_\text{trap}^2}\biggr)^{\ell-1}\\
\times\exp\Bigl\{-
(y^2-2xy)/2\sigma_\text{ou}^2\Bigr\}.
\end{multline}

After identifying the factor
\begin{equation}
\exp\Bigl\{-
\bigl(y^2-2xy\bigr)/2\sigma_\text{ou}^2\Bigr\}    
\end{equation}
with the generating function of the Hermite polynomials $H_n(x)$ (here we resort to the probabilist definition), i.e.,
\begin{equation}
e^{- 
\bigl(y^2-2xy\bigr)/2\sigma_\text{ou}^2}=\sum_{n=0}^\infty \frac{1}{n!}H_n\biggl(
\frac{x}{\sigma_\text{ou}}\biggr)\biggl(
\frac{y}{\sigma_\text{ou}}\biggr)^n,
\end{equation}
an carrying out the integral we can write $P_\text{st}(x)$ as
\begin{multline}
P_\text{st}(x)=G_\text{st}(x)\frac{\Gamma\bigl(\ell+\frac{1}{2}\bigr)}{\sqrt{\pi}}\sum_{n=0}^{\infty}\frac{\Gamma\bigl(n+\frac{1}{2}\bigr)}{(2n)!\Gamma\bigl(n+\ell+\frac{1}{2}\bigr)}\\
\times H_{2n}\biggl(
\frac{x}{\sigma_\text{ou}}\biggr)\biggl(\frac{v^2}{\mu\kappa D_T}\biggr)^{n},
\label{StationaryDist-Hermite}
\end{multline}
where $\sqrt{D_T\mu\kappa}$ measures the average particle speed imparted by thermal fluctuations. This expression for the stationary distribution of the position depends explicitly on the parameters $\ell$ and $\mathcal{D}^{-1}=v^2/(\mu\kappa D_T)$ and factorizes into the product of the equilibrium distribution $G_\text{st}(x)$ and a function that modulates it carrying the non-equilibrium effects of active motion.
\begin{figure*}
\includegraphics[width=0.24\textwidth]{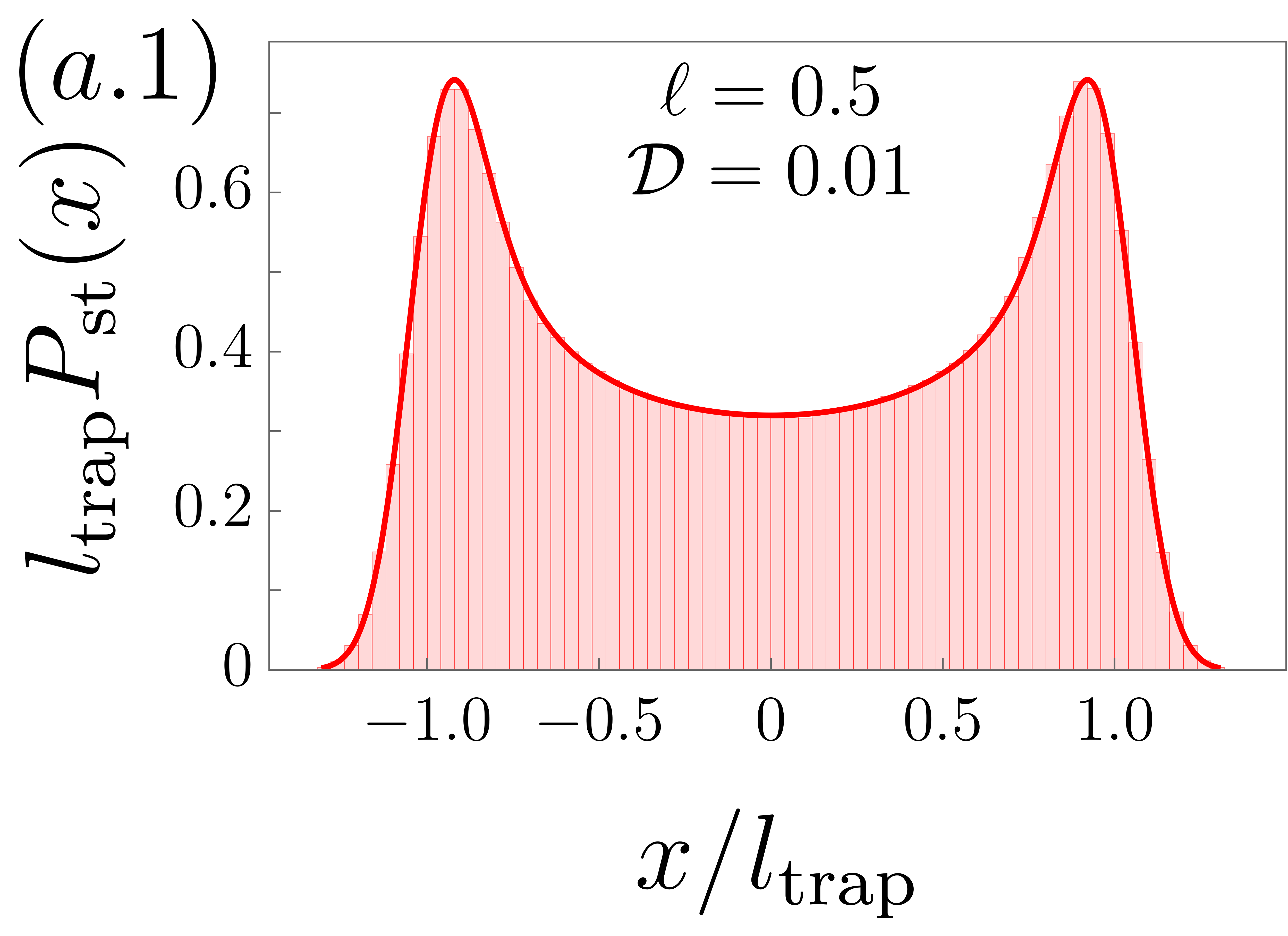}
\includegraphics[width=0.24\textwidth]{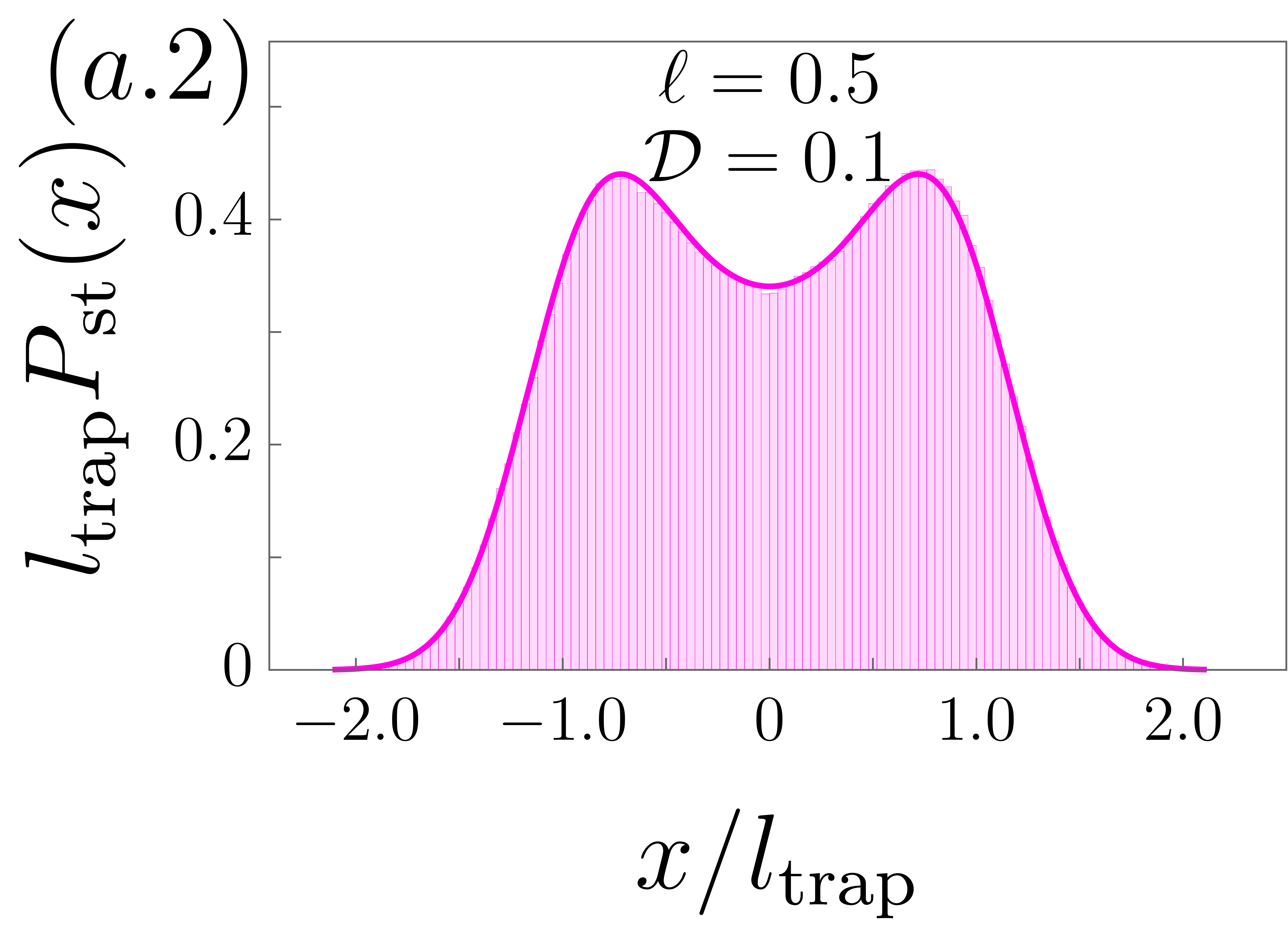}
\includegraphics[width=0.24\textwidth]{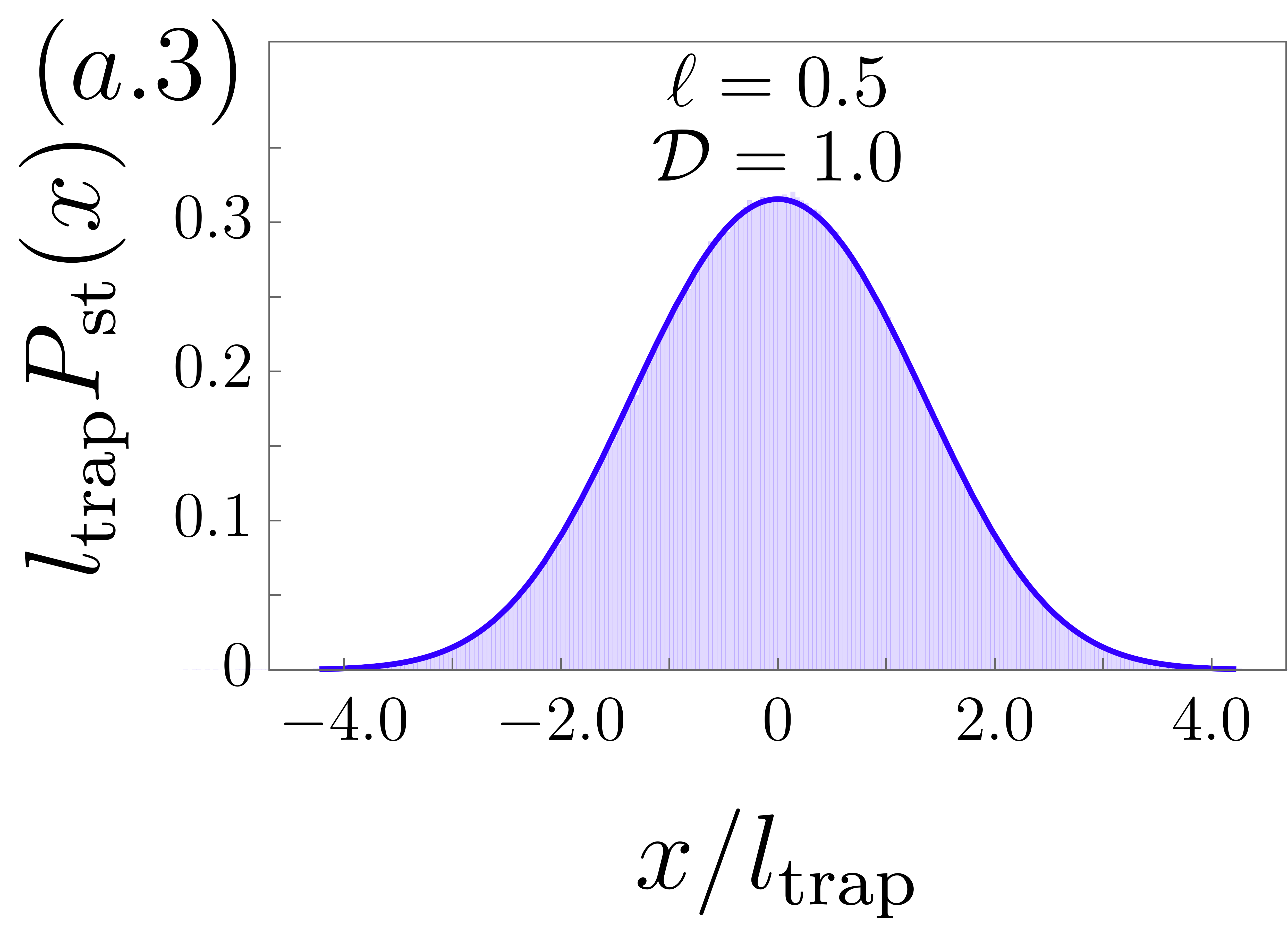}
\includegraphics[width=0.25\textwidth]{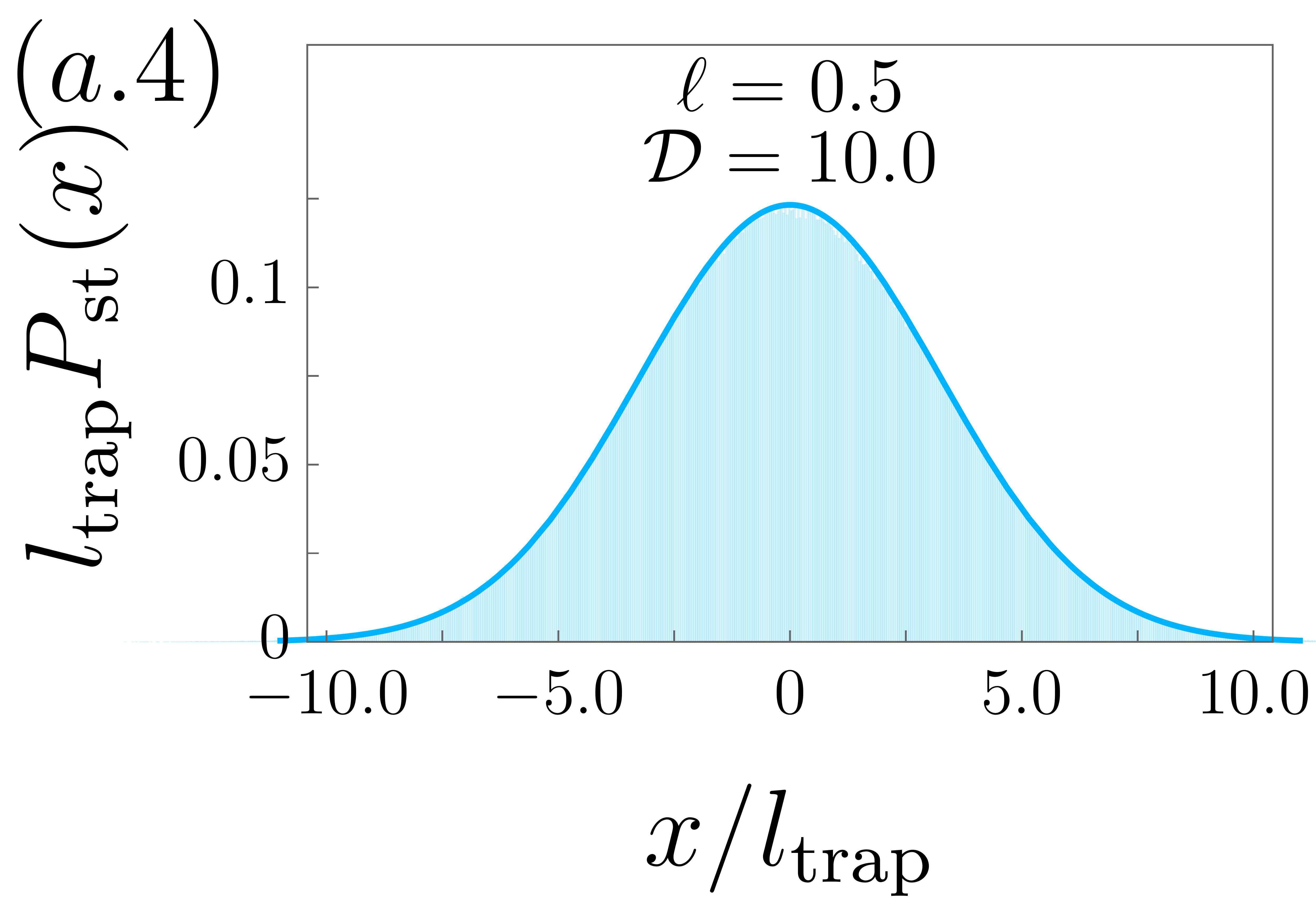}\\
\includegraphics[width=0.24\textwidth]{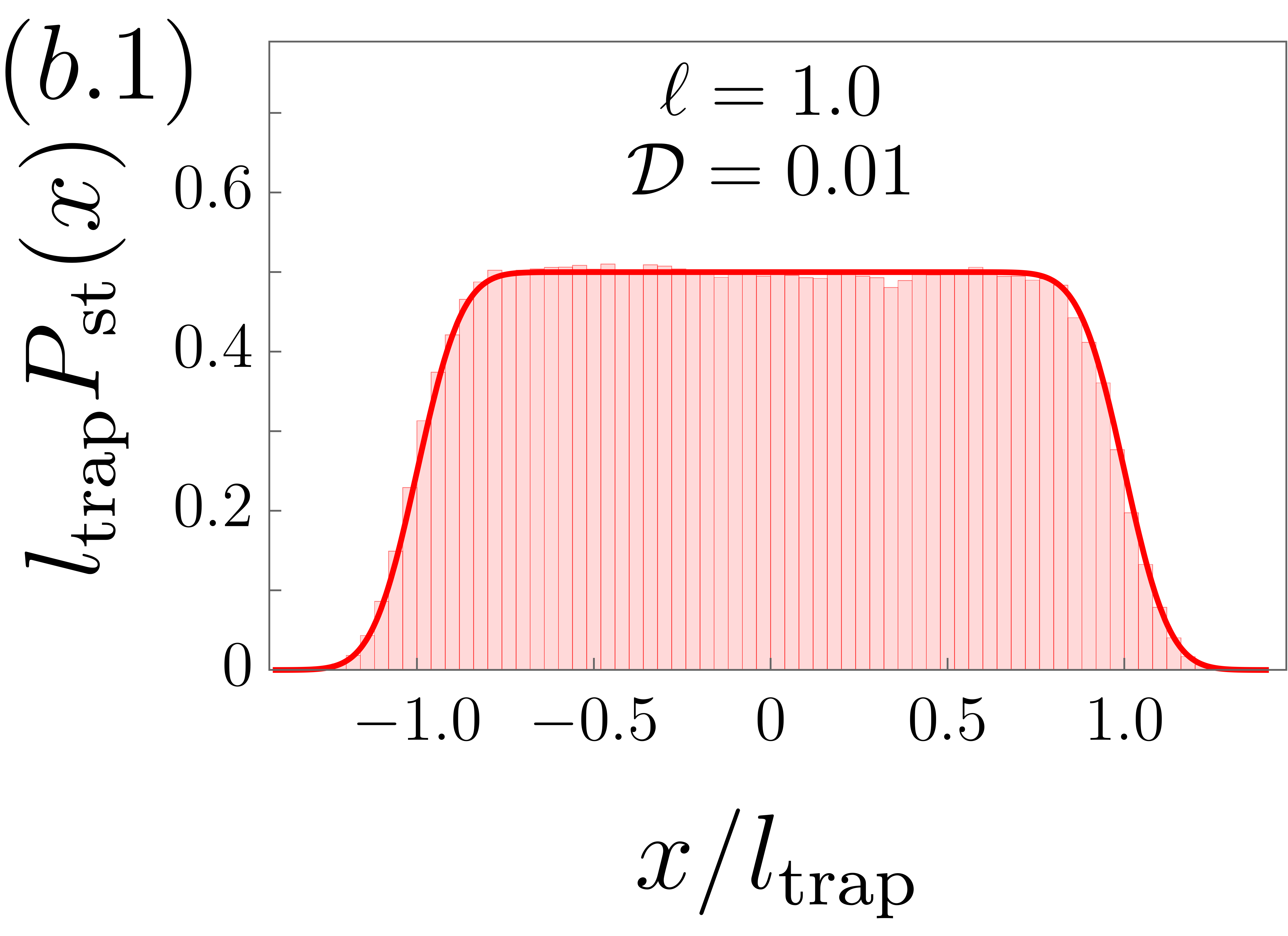}
\includegraphics[width=0.24\textwidth]{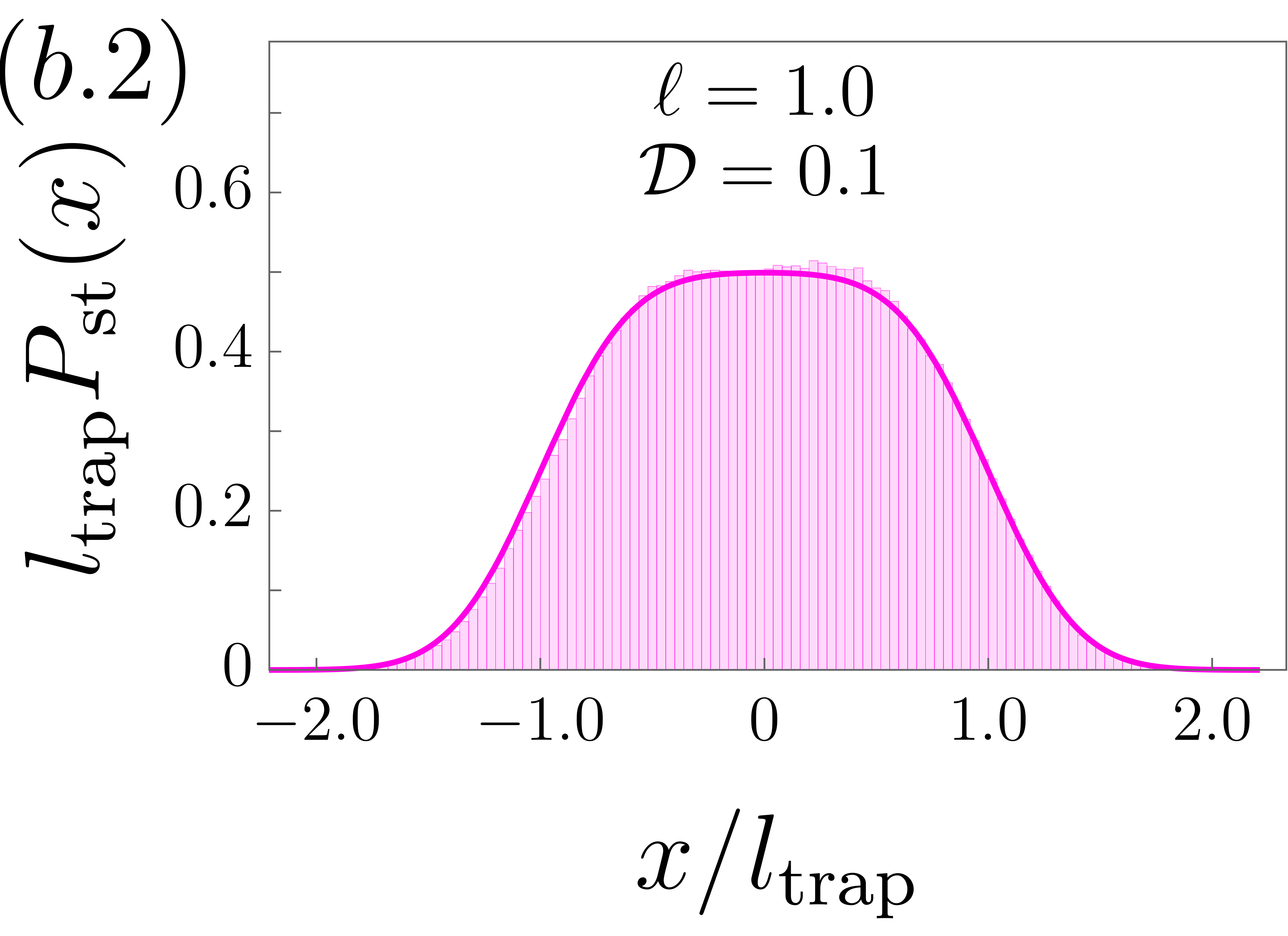}
\includegraphics[width=0.24\textwidth]{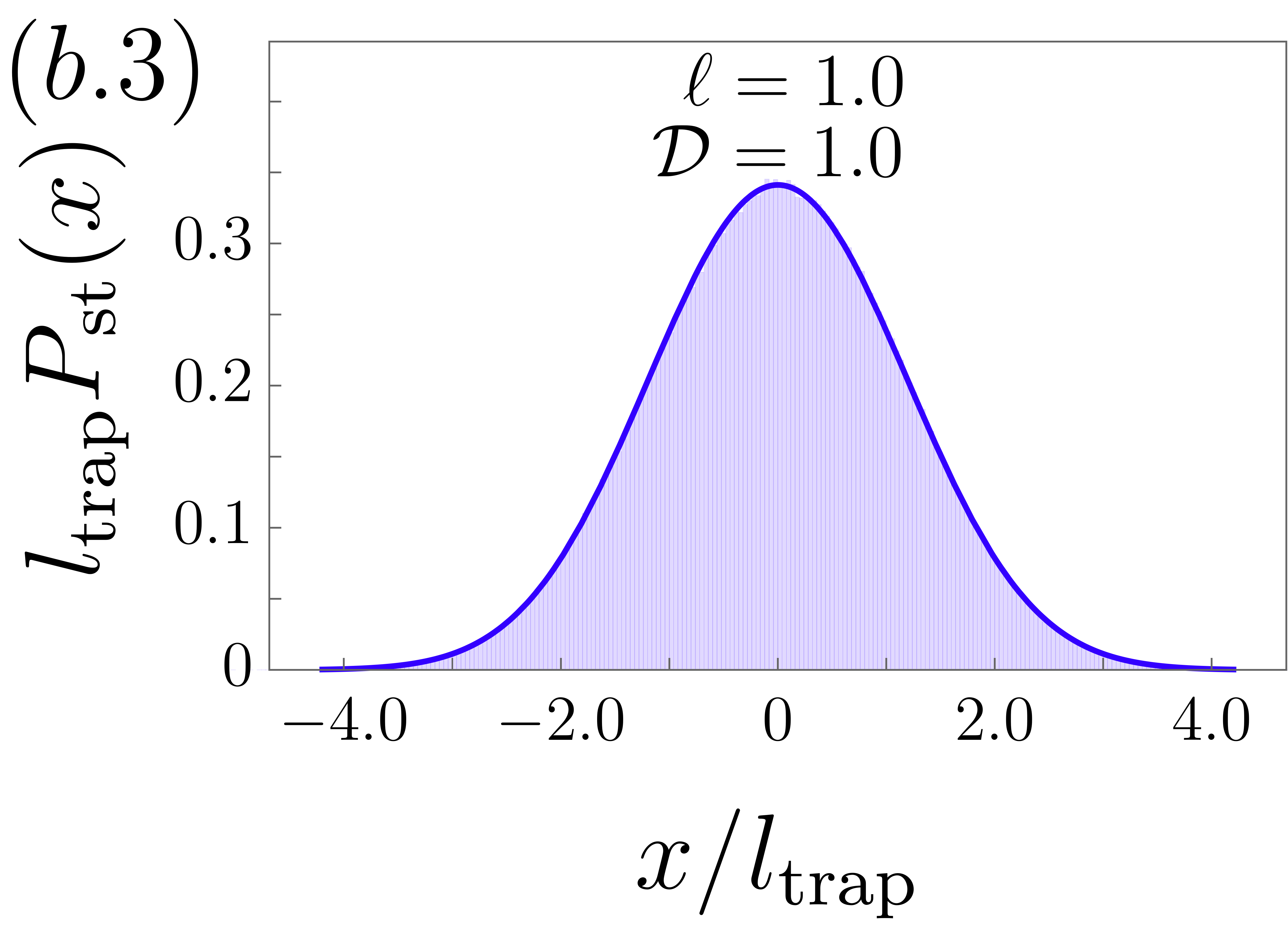}
\includegraphics[width=0.245\textwidth]{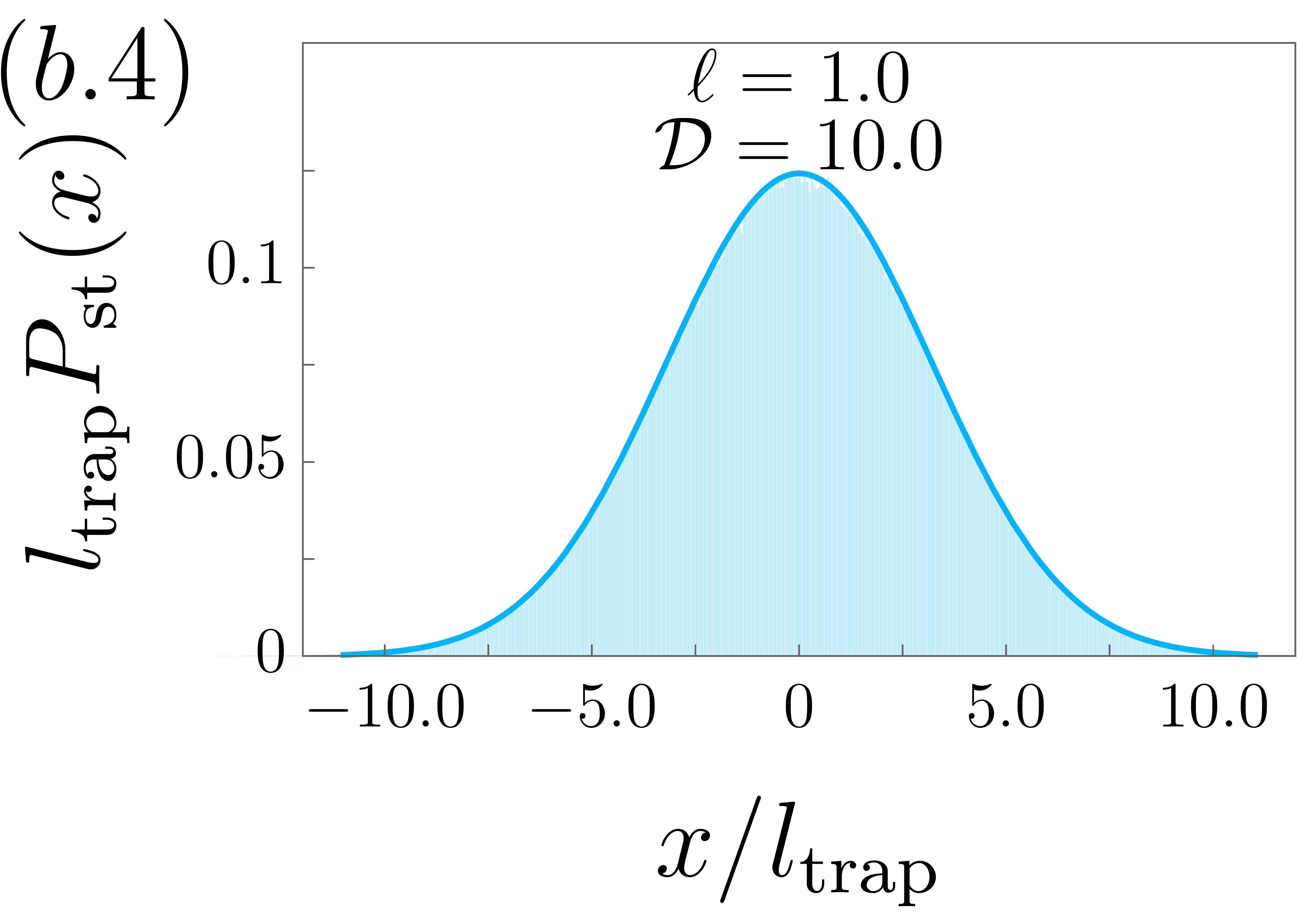}\\
\includegraphics[width=0.24\textwidth]{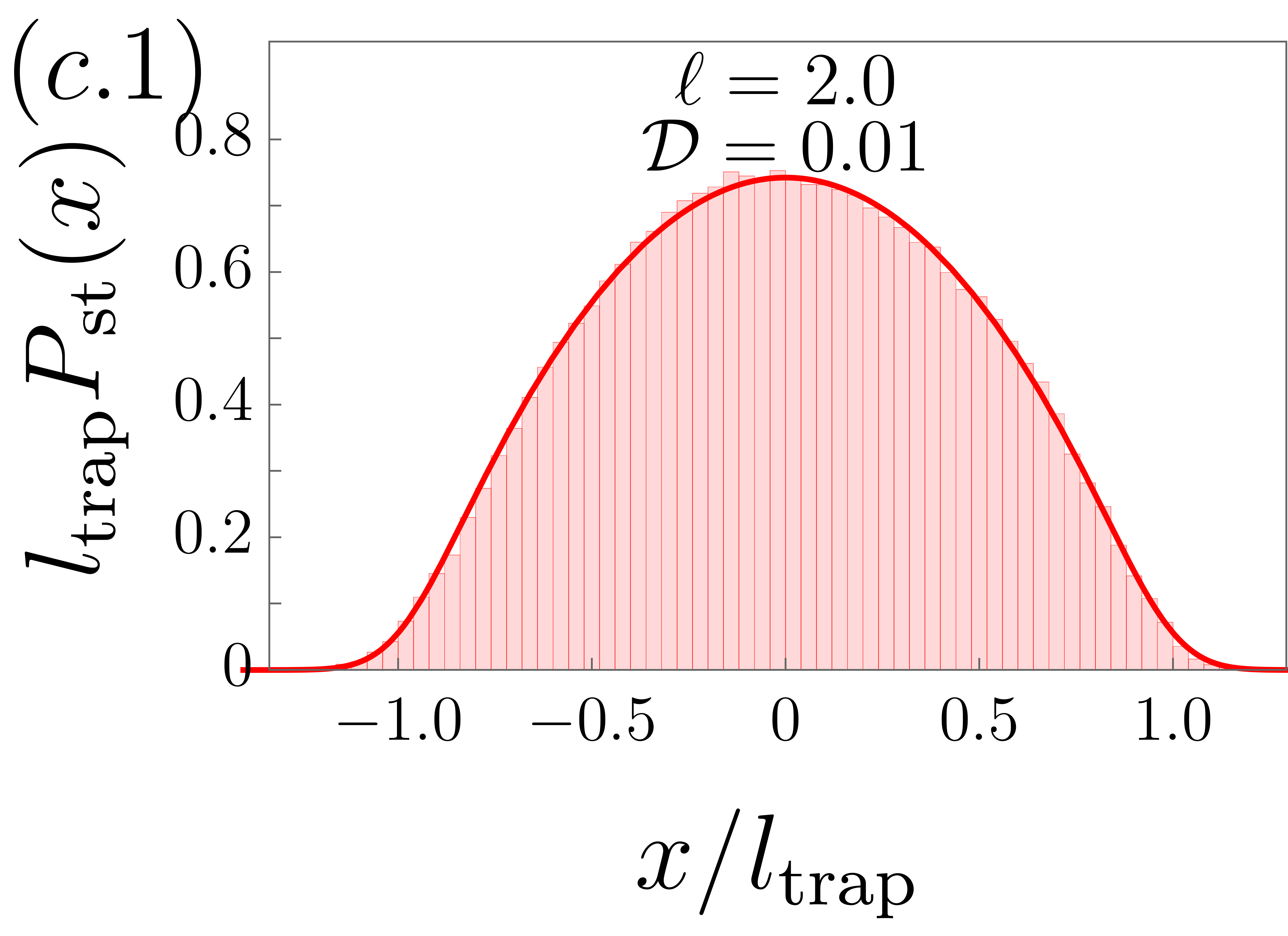}
\includegraphics[width=0.24\textwidth]{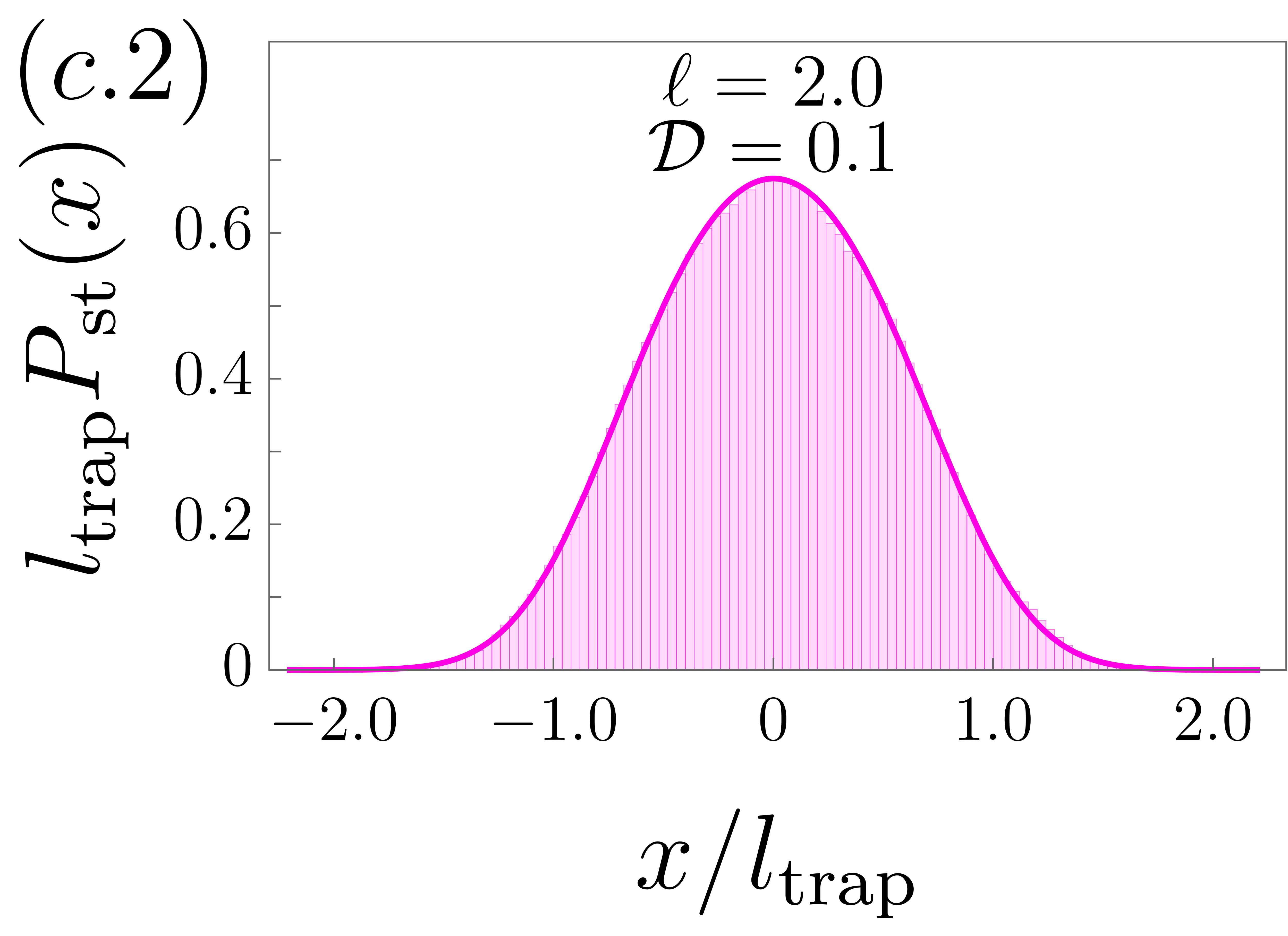}
\includegraphics[width=0.24\textwidth]{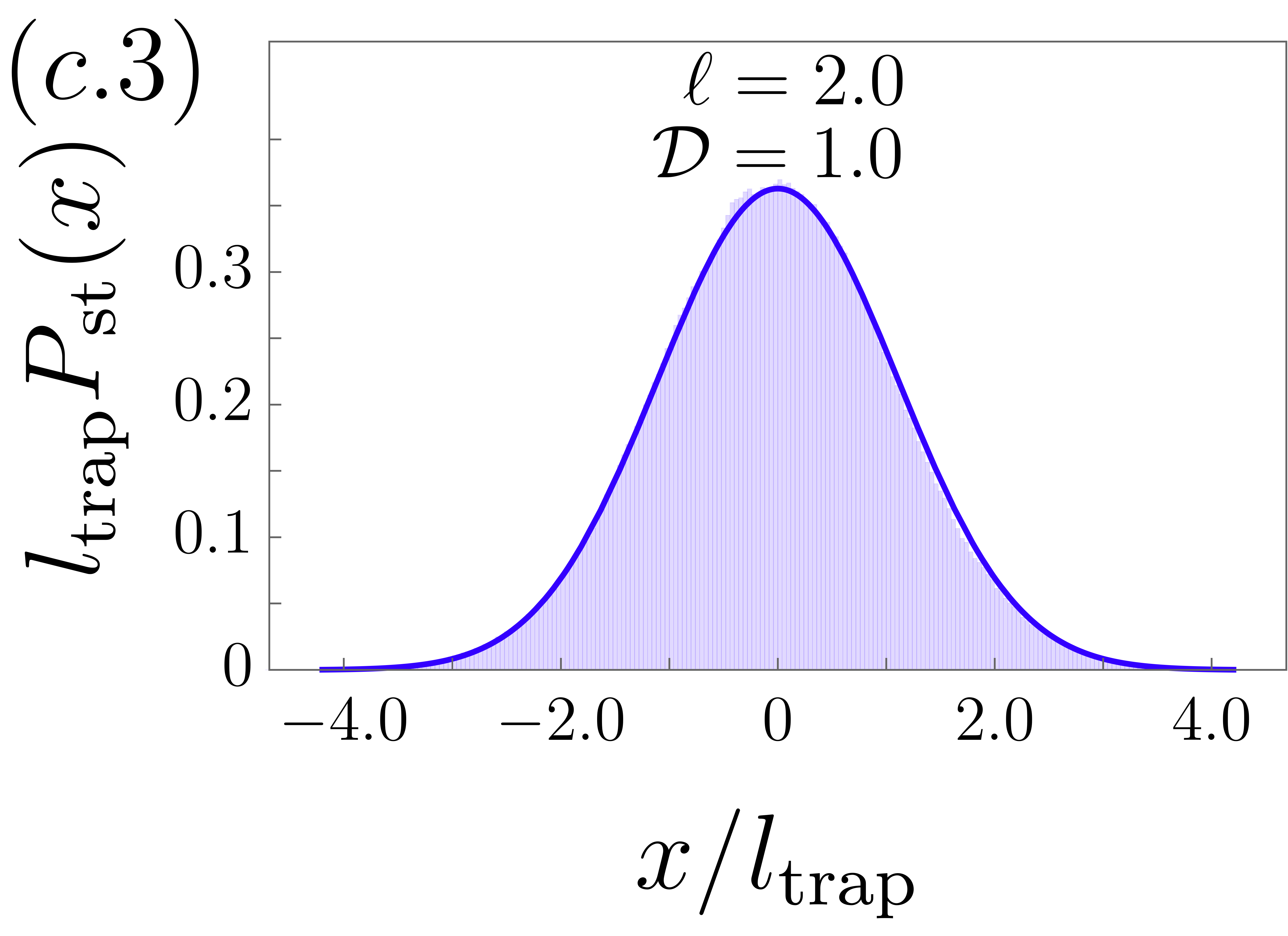}
\includegraphics[width=0.245\textwidth]{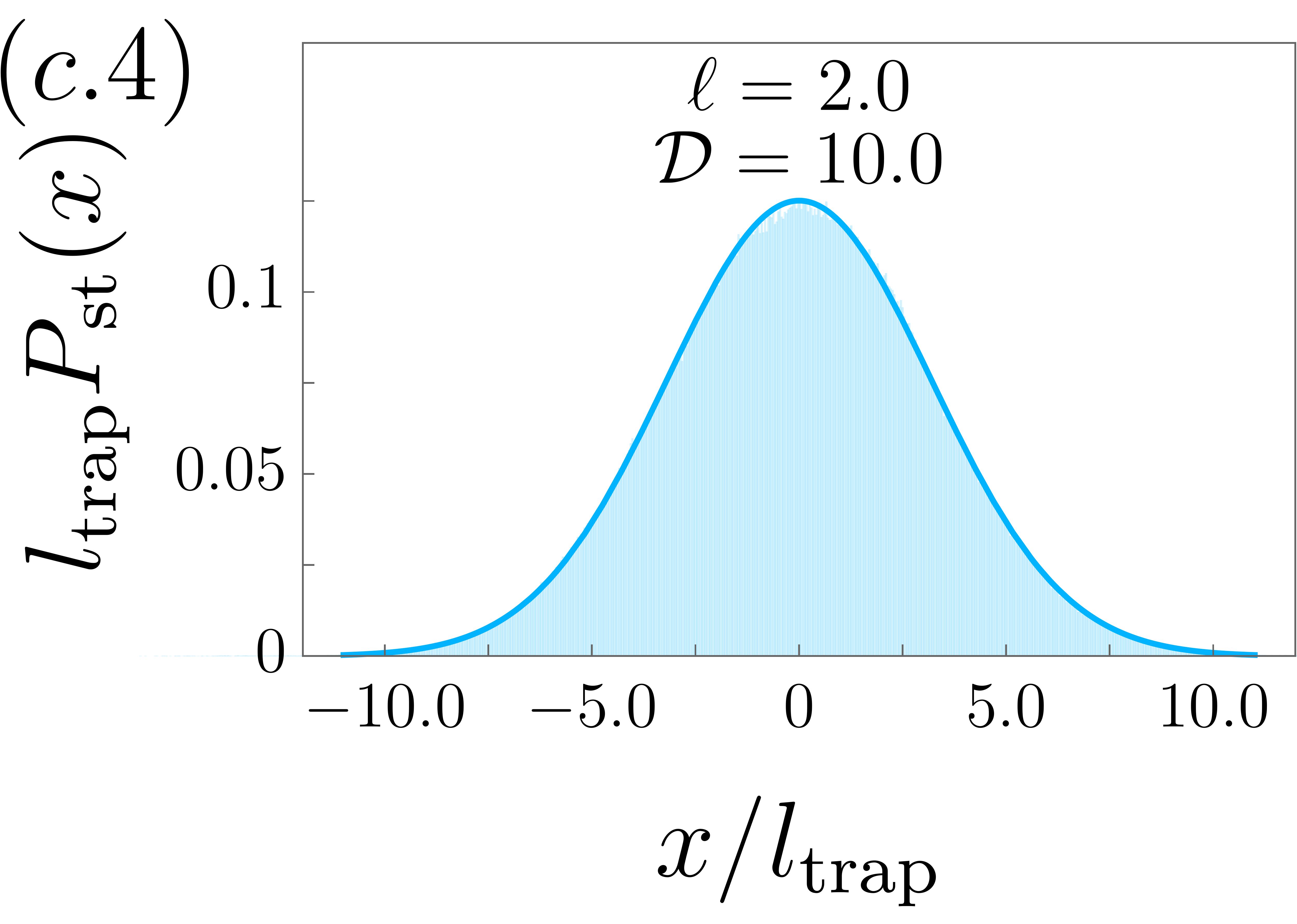}
\caption{\label{fig:StationaryDistributions}Dimensionless stationary distribution $l_\text{trap}P_\text{st}(x)$ of the particle positions in the harmonic potential $\frac{1}{2}\kappa x^2$, as function of the dimensionless position $x/l_\text{trap}$ for different values of $\ell=l_\text{trap}/l_\text{pers}$, from right to left:  0.5 row ($a$), 1.0 row ($b$) and 2.0 row ($c$); and different values of $\mathcal{D}=D_T(\mu\kappa/v^2)$, from top to bottom: 0.01 column (1), 0.1 column (2), 1.0 column (c) and 10.0 column (4). $l_\text{trap}=v/\mu\kappa$ and $l_\text{pers}=v(2/\alpha)$ are the characteristic trapping and persistent lengths respectively, and $\mu$ the mobility.}
\end{figure*}
Notice that by the Legendre duplication formula $\Gamma\bigl(m+\frac{1}{2}\bigr)=\sqrt{\pi}(2m)!/(2^{2m}m!)$, we have that
\begin{subequations}
\begin{equation}\label{factor}
 \frac{1}{\sqrt{\pi}}\frac{\Gamma\bigl(\ell+\frac{1}{2}\bigr) \Gamma\bigl(n+\frac{1}{2}\bigr)}{(2n)!\Gamma\bigl(n+\ell+\frac{1}{2}\bigr)}=\frac{1}{2^{2n}n!}\frac{\Gamma\bigl(\ell+\frac{1}{2}\bigr)}{\Gamma\bigl(n+\ell+\frac{1}{2}\bigr)}
\end{equation}
and since
\begin{equation*}
 \frac{\Gamma\bigl(\ell+\frac{1}{2}\bigr)}{\Gamma\bigl(n+\ell+\frac{1}{2}\bigr)}=\frac{1}{\bigl(\ell + \frac{1}{2}\bigr) \bigl(\ell + \frac{1}{2} +1\bigr) \cdots \bigl(\ell+n -\frac{1}{2}\bigr)}
\end{equation*}
the right-hand side of Eq.~\eqref{factor} can be written as
\begin{multline}
\frac{1}{[2(2\ell+1)][4(2\ell+3)]\cdots[2n(2l+2n-1)]}=\\
\prod_{j=1}^{2n} \frac{1}{j+ 2\ell p_j}
\end{multline}
where $p_j = 1$ if $j$ odd and $p_j = 0$ otherwise.
\end{subequations}

With this last result, Eq.~\eqref{StationaryDist-Hermite} can be written as
\begin{multline}\label{eq:dist_st_per1}
P_{\text{st}} (x) = G_\text{st}(x) \biggl[1+ \sum_{n=1}^{\infty}
\biggl(\frac{v^2}{\mu \kappa D_T}\biggr)^{n}\times\\ 
H_{2n}\left( \sqrt{\frac{\mu \kappa}{D_T}} x \right) \prod_{j=1}^{2n} \frac{1}{j+ 2\ell p_j}  \biggr],
\end{multline}
which is exactly the result obtained in \cite{Garcia-MillanJStatMech2021} using field theory methods.

The Fourier transform of Eq.~\eqref{StationaryDist} can be carried out explicitly, given by
\begin{equation}
\label{FT-StationaryDist}
\hat{P}_\text{st}(k)={_0F}_1\biggl(;\ell+\frac{1}{2},-\frac{(l_\text{trap}k)^2}{4}\biggr)\cdot e^{-\mathcal{D}(l_\text{trap}k)^2/2},
\end{equation}
where ${_0F}_1\bigl(;\nu,z\bigr)=\sum_{n=0}^\infty z^n/n!(\nu)_n$, $(\nu)_n$ is a standard hypergeometric function, $(\nu)_n=\nu(\nu+1)\ldots(\nu+n-1)$ being the Pochhammer symbol, and we have used that $\mathcal{D}=\sigma_{ou}^2/l_\text{trap}^2$.

\paragraph{The diffusive limit.-} In the limit, $\ell\rightarrow\infty$, the persistence length is negligible with respect to the trapping length, this occurs when both, $\alpha$ and $v$, goes to $\infty$, but with finite  $D_\text{act}\equiv v^2/\alpha$, thus $l_\text{pers}=v/(\alpha/2)\rightarrow0$. After rearranging Eq.~\eqref{EqStatDist}, it can be shown straightforwardly that in this limit the position distribution of the active part, $p_\text{st}(x)$, can be written as
\begin{equation}
\frac{\partial}{\partial x}p_\text{st}(x)=-\frac{\alpha}{2v^2}\mu\kappa\, 2x\,p_\text{st}(x),
\end{equation}
which leads, after normalization, to the thermal Gaussian stationary distribution
\begin{equation}\label{BoltzmannHO}
p_\text{st}(x)=\sqrt{\frac{\mu\kappa}{2\pi D_\text{act}}}\exp\left\{-\frac{\mu\kappa}{2D_\text{act}}x^{2}\right\},
\end{equation}
but with diffusion constant $D_\text{act}.$
With this result the total stationary distribution in the regime $\ell\rightarrow\infty$ as
\begin{equation}
\label{DiffLimit}
P_{st}(x)=\sqrt{\frac{\mu\kappa}{2\pi D_\text{eff}}}\exp\left\{-\frac{\mu\kappa}{2D_\text{eff}}x^{2}\right\}
\end{equation}
with the effective diffusion coefficient $D_\text{eff}=D_T+D_\text{act}$. This result can also be derived from Eq.~\eqref{FT-StationaryDist} by noticing that in the limit $\ell\rightarrow\infty$, we have ${_0F}_1(;\ell,-(l_\text{trap}k)^2/4)\sim e^{-k^2 l_\text{trap}l_\text{pers}/4}$, thus Fourier inversion of $\hat{P}_\text{st}(x)$ leads to \eqref{DiffLimit}.

\paragraph{The persistent regime.-} In the case $\ell\rightarrow0$, persistence makes the particles to dwell around the positions $\pm l_\text{trap}$, which is a well-known effect of active motion that emerges as consequence of the strong correlations induced by persistent motion. In this limit we have  ${_0F}_1\bigl(;1/2,-(l_\text{trap}k)^2/4\bigr)=\cos(kl_\text{trap})$, Fourier inversion of the resulting $\hat{P}_\text{st}(k)$ leads to the overlapping Gaussian distributions $G_\text{st}(x)$ given in \eqref{Gaussian} centered at $\pm l_\text{trap}$, namely 
\begin{equation}
\label{PersStatDistribution}
P_\text{st}(x)=\frac{1}{2}[G_\text{st}(x+l_\text{trap})+G_\text{st}(x-l_\text{trap})].
\end{equation}
As thermal fluctuations diminish, the effects of active motion intensifies, consequently, the overlapping between the two contribution in \eqref{PersStatDistribution} diminishes.

In Fig.~\ref{fig:StationaryDistributions}, the analytical expression for $P_\text{st}(x)$ given in Eq.~\eqref{StationaryDist} (solid lines), is compared with the corresponding distribution of positions obtained from the numerical integration of the stochastic differential equation \eqref{Langevin}
(histogram bars). Particularly, we show the transition of $P_\text{st}(x)$ from
the persistent regime, $\mathcal{D}=0.01$ (column 1), to the diffusive one, $\mathcal{D}=10.0$ (column 4), passing through the intermediate values 0.1 (column 2) and 1.0 (column 3) for $\ell=0.5$ (row $a$), 1.0 (row $b$) and 2.0 (row $c$). In each case of fix $\ell$, as temperature increases, the features of the persistent regime fades towards a distribution of the positions with a maximum at the enter of the trap (weak non Gaussianity is observed). In the persistent regime (column 1), a visit depletion of the trap center is observed whenever $l_\text{pers}>l_\text{trap}$ (see ($a.1$) for $l_\text{pers}=2\, l_\text{trap}$), exhibiting bimodality around $\pm l_\text{trap}$. This effect vanishes for $l_\text{pers}\le l_\text{trap}$ as shown in panels ($b.1$) and ($c.1$), exhibiting strongly non Gaussianity. 

After use of the relation between ${_0F}_1(;\nu,z)$ and the Bessel functions $J_\mu(z)$, namely 
$${_0F}_1\Bigl(;\frac{1}{2}+\ell,-\frac{k^2}{4}\Bigr)=\Gamma\Bigl(\frac{1}{2}+\ell\Bigr)\frac{|k|^{1/2-\ell}}{2^{1/2-\ell}}J_{\ell-1/2}(k),$$
we are able to provide analytical expressions for the ISF for some particular values of $\ell$. In the persistent regime, $\ell<1$, we have
\begin{subequations}
\label{ISF-ParticularCases}
\begin{equation}
\hat{P}_\text{st}(k)=J_0(k\,l_{\mathrm{trap}})\, e^{-\mathcal{D}(l_{\mathrm{trap}}k)^{2}/2}
\end{equation}
for $\ell=\tfrac{1}{2}$. At threshold $\ell=1$ we have
\begin{equation}
\label{ell=1}
\hat{P}_\text{st}(k)= \frac{\sin(kl_\text{trap})}{kl_\text{trap}} e^{-\mathcal{D}(l_{\mathrm{trap}}k)^{2}/2},
\end{equation}
while for $\ell=2$ we get
\begin{multline}
\label{ell=2}
\hat{P}_\text{st}(k)= \frac{3}{k^2l_\text{trap}^2}\biggl(\frac{\sin(kl_\text{trap})}{kl_\text{trap}}-\cos(kl_\text{trap})\biggr)\\
\times e^{-\mathcal{D}(l_{\mathrm{trap}}k)^{2}/2}.
\end{multline}
\end{subequations}
The inverse Fourier transform of case $\ell=1$ can be evaluated exactly as
\begin{equation}
\label{TF_StatDistribution_l1}
  P_\text{st}(x)=\frac{1}{4l_{\mathrm{trap}}}
  \left[\operatorname{erf}\!\left(\frac{x+l_{\mathrm{trap}}}{\sqrt{2}\,\sigma_\text{ou}}\right)
  -\operatorname{erf}\!\left(\frac{x-l_{\mathrm{trap}}}{\sqrt{2}\,\sigma_\text{ou}}\right)\right],
\end{equation}
since the inverse Fourier transform of $\sin(kl_\text{traps})/kl_\text{trap}$ leads to the rectangle function $\theta (l_\text{trap}-|x|)/l_\text{trap}$ as can be seen from Eq.~\eqref{p-act-st}.

The special cases listed above are not independent. By use of the identity $\tfrac{d}{dz}\bigl[z^{-\nu}J_{\nu}(z)\bigr]=-z^{-\nu}J_{\nu+1}(z)$, we have that for consecutive values of $\ell$ the following relation holds
\begin{equation}
\label{eq:ladder_fou}
  \hat P_{\text{st},(\ell+1)}(k) = -(2\ell+1)
  \left( \frac{1}{l_{\mathrm{trap}}^{2}\,k}\,\frac{d}{dk}
  + \mathcal{D} \right) \hat P_{\text{st},(\ell)}(k),
\end{equation}
from which $\hat P_{\text{st},(2)}(k)$ and $\hat P_{\text{st},(1)}(k)$ are recovered, and by iteration, the entire family of integer values of $\ell$ can be generated. The additive term $\mathcal{D}$ compensates the $\ell$-independent Gaussian factor. Notice that Eq.~\eqref{eq:ladder_fou} shifts $\ell$ by one unit, so the half-integer sequence starting at $\hat P_{\text{st},(1/2)}(k)$ forms a separate sequence of stationary distributions for which no differential operator connects it with the sequence of integers.

\subsection{The fluctuations of the energy and the distance from nonequilibrium}

In the stationary regime, the run-and-tumble contribution to the particle motion induces the deviation from the Gaussian distributions characteristic of passive Brownian motion. A standard measure of non Gaussianity is the kurtosis $\varkappa(p)=\bigl\langle(x-\langle x\rangle)^4\bigr\rangle/\bigl\langle(x-\langle x\rangle)^2\bigr\rangle^2$, with $\langle g(x)\rangle=\int dx\, g(x)p(x)$. On the other hand, energy fluctuations for an arbitrary trapping potential $U(x)$ are given by 
\begin{equation}
\Delta E=\Bigl\langle\bigl[U(x)-\langle U(x)\rangle\bigr]^{2}\Bigr\rangle^{1/2}.
\end{equation}
For the case under consideration here, $U(x)=\frac{1}{2}\kappa x^2$, and thus the energy fluctuations are 
\begin{subequations}
\label{EnergyFluct}
\begin{align}
\Delta E&=\frac{1}{2}\kappa\Bigl[\bigl\langle x^{4}\bigr\rangle_{P_\text{st}}-\bigl\langle x^{2}\bigr\rangle_{P_\text{st}}^{2}\Bigr]^{1/2}\\
&=E\cdot\Bigl(\varkappa(P_\text{st})-1\Bigr)^{1/2},
\end{align}
where we have used that $\langle x\rangle_{P_\text{st}}=0$, $E$ is the average energy (see appendix \ref{sect:moments})
\begin{align}
E=&\frac{1}{2}\kappa \bigl\langle x^{2}\bigr\rangle_{P_\text{st}}\\
=&\frac{1}{2}\kappa l_\text{trap}^2\biggl(\mathcal{D}+\frac{1}{4(\ell+\frac{1}{2})}\biggr),
\end{align}
and $\varkappa(P_\text{st})$ denotes the kurtosis of $P_\text{st}(x)$,
\end{subequations}
which for Gaussian distributions attains the invariant value $3$ under translations of the mean and under the scale transformation of the variance. The exact expression for $\varkappa(P_\text{st})$ is given by (see appendix \ref{sect:moments})
\begin{equation}
\varkappa(P_\text{st})=
\frac{3\mathcal{D}^2+\dfrac{3}{2}\dfrac{\mathcal{D}}{\ell+\frac{1}{2}}+\dfrac{3}{16}\dfrac{1}{(\ell+\frac{1}{2})(\ell+\frac{3}{2})}}{\biggl[\mathcal{D}+\dfrac{1}{4(\ell+\frac{1}{2})}\biggr]^2}.
\end{equation}

\begin{figure}
\includegraphics[width=\columnwidth]{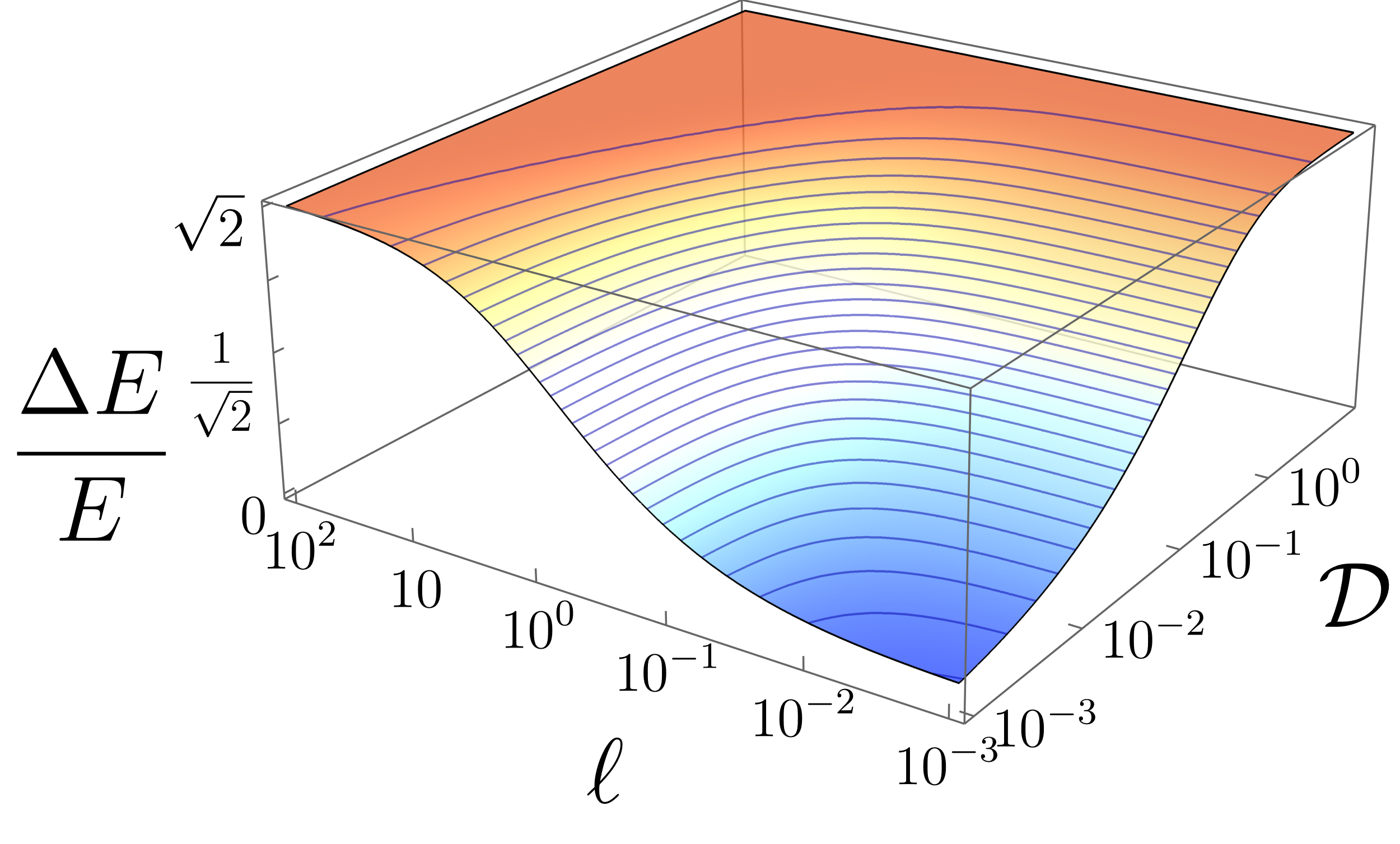}
\caption{The ratio of the energy fluctuations to the mean energy \eqref{Ratio-SD2E} is shown as function of  $\ell=l_\text{trap}/l_\text{pers}$ and $\mathcal{D}=D_T(\mu\kappa/v^2)$.}
\label{Ratio-SD2E}
\end{figure}
From Eqs.~\eqref{EnergyFluct}, the ratio of the standard deviation to the mean of the energy is given by
\begin{equation}
\frac{\Delta E}{E}=\Bigl(\varkappa(P_\text{st})-1\Bigr)^{1/2},
\end{equation}
which for the Gaussian equilibrium distribution of Boltzmann attains the value $\sqrt{2}$, reached when $\ell$, $\mathcal{D}$, or both tend to $\infty$. This ratio diminishes as active motion become greater, either by $\ell\rightarrow0$, or $\mathcal{D}\rightarrow0$, indicating that the particle can have very large positional fluctuations while on the contrary having almost no energy fluctuations. In Fig.~\ref{Ratio-SD2E} $\Delta E/E$ is shown as function of $\ell$ and $\mathcal{D}$, the countour lines for constant $\Delta E /E=\text{constant}$ are indicated. 

\begin{figure}
\includegraphics[width=\columnwidth]{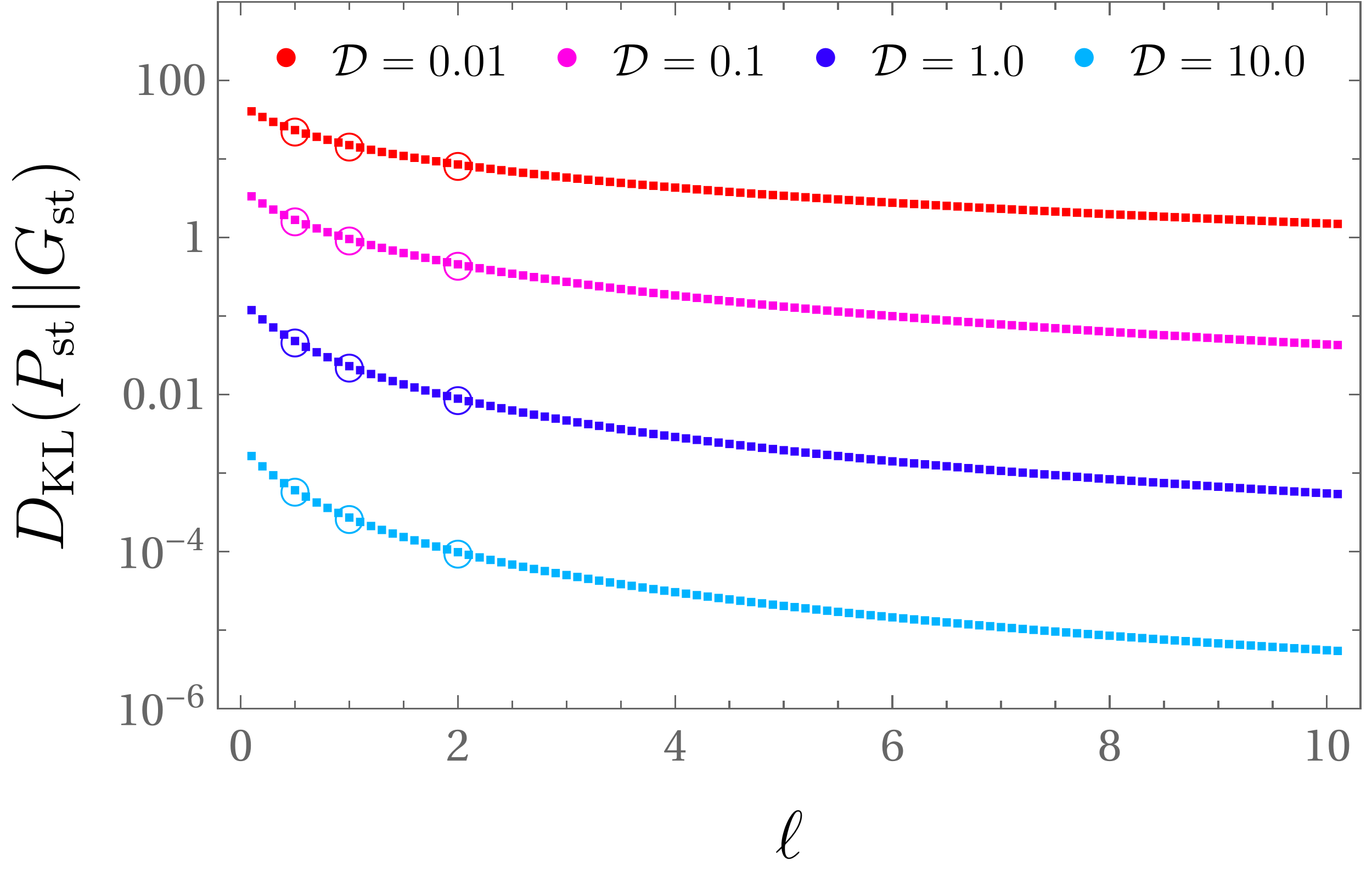}
\caption{Kullback-Leibler distance between $P_\text{st}$ and $G_\text{st}$, $D_\text{KL}(P_\text{st}||G_\text{st})$, as function of $\ell=l_\text{trap}/l_\text{pers}$ for different values of $\mathcal{D}=\sigma_\text{ou}^2/l_\text{trap}^2$, namely 0.01 (red), 0.1 (magenta), 1.0 (purple) and 10.0 (cyan), the empty circles mark the values of $\ell$ as in Fig.~\ref{fig:StationaryDistributions}. }
\label{fig:KL}
\end{figure}

Finally, notice that in Eqs.~\eqref{StationaryDist}, \eqref{StationaryDist-Hermite} and \eqref{eq:dist_st_per1}, a factor carrying the effects of run-and-tumble motion and that modulates the equilibrium distribution $G_\text{st}(x)$ can be identified. This factor determines the non-equilibrium nature of $P_\text{st}(x)$ induced by active motion, and its effects can be quantified by the Kullback-Leibler distance 
\begin{equation}
D_\text{KL}(P_\text{st}||G_\text{st})=\int_{-\infty}^\infty dx\, P_\text{st}(x)\ln\frac{P_\text{st}(x)}{G_\text{st}(x)}.
\end{equation}
In Fig.~\ref{fig:KL} the dependence of $D_\text{KL}$ on $\ell$ is shown for different values od $\mathcal{D},$ 0.01 (red), 0.1 (magenta), 1.0 (purple) and 1.0 (cyan). As expected, small thermal noise corresponds to the case farthest case from equilibrium since particle motion is dominated by the run-and-tumble process (see $\mathcal{D}=0.01$), and the convergence to equilibrium occurs as $\mathcal{D}$ increases. In all the cases shown in Fig.~\ref{fig:StationaryDistributions} the non-equilibrium nature of the distribution is appreciated (marked with empty circles in Fig.~\ref{fig:KL}). As $\ell$ increases, $D_\text{KL}$ diminishes monotonically, this is clearly understood since as the persistence length becomes smaller than the trapping length, the effects of run-and-tumble fade.

\section{\label{sect:Conclusions}Final comments and concluding remarks}

We have presented an exact analysis of the stationary statistics of one-dimensional run-and-tumble particles confined by a harmonic potential and subject to thermal fluctuations from an equilibrium bath. By exploiting the fact that the harmonic trap couples the active and thermal degrees of freedom through the Ornstein-Uhlenbeck propagator, we established that the full stationary distribution factorizes, in the sense of a convolution, into the Ornstein-Uhlenbeck distribution and the stationary distribution of the purely athermal run-and-tumble problem, Eq.~\eqref{TotalP}. This route is conceptually transparent and requires no auxiliary field-theoretic construction, yet it reproduces and, in the appropriate limit, corrects results previously obtained by more elaborate means, including the identification of the correct propagator that must enter the convolution, namely, the Ornstein-Uhlenbeck kernel rather than the free-diffusion one used in an earlier work. The validity of the convolution was confirmed independently in Fourier space and through direct numerical integration of the Langevin dynamics.

The construction presented, yields closed-form expressions for the stationary distribution Eq.~\eqref{StationaryDist}, from which the expression in terms of Hermite-polynomial series in Eq.~\eqref{eq:dist_st_per1} \cite{Garcia-MillanJStatMech2021} is recovered. For special values of the ratio between the trapping and persistence lengths, $\ell$, explicit elementary functions are shown in Eqs.~\eqref{ISF-ParticularCases} for the intermediate scattering function, while the explicit dependence on space coordinate  is possible for $\ell\rightarrow0$, Eq.~\eqref{PersStatDistribution} and $\ell=1$, Eq.~\eqref{TF_StatDistribution_l1}. The two dimensionless control parameters of the problem, $\ell$ and $\mathcal{D}$, fully determine the crossover between a strongly non-Gaussian, boundary-peaked distribution characteristic of persistent active motion and the Gaussian, Boltzmann-like distribution recovered in the diffusive limit.

The nonequilibrium character of the confined active particle is further quantified through the energy fluctuations and the associated ratio $\Delta E/E$, which departs markedly from the equilibrium value $\sqrt{2}$ as persistence dominates over trapping or thermal fluctuations, indicating that the particle can sustain large positional excursions while its energy remains comparatively well defined. This behavior, together with the explicit spatial dependence of the effective bath implied by our results, provides a concrete, analytically accessible measure of how far the trapped active system departs from equilibrium.

More broadly, the approach developed here, treating the thermal and active contributions as convolved rather than additive processes, offers a general and technically economical strategy for incorporating thermal noise into confined active-matter problems. It should extend naturally to other confining potentials, to higher-dimensional geometries, and to related classes of active-particle models, and it connects directly to observables such as the intermediate scattering function that are accessible in dynamic light scattering and differential dynamic microscopy experiments, opening a promising avenue for the direct experimental characterization of the effective, spatially heterogeneous bath experienced by confined active particles.

\appendix
\section{The convolution solution}
\label{Proof}
An alternative way to proof the solution \eqref{TotalP} is presented in this appendix. We start with the total probability density $P(x,t)$ given by $P(x,t)=\bigl\llangle\delta\bigl(x-x(t)\bigr)\bigr\rrangle_{\sigma,\xi}$, where $\llangle\cdot\rrangle_{\sigma,\xi}$ denotes the expectations value over realizations of the stochastic processes $\sigma(t)$ and $\xi(t)$, $x(t)$ being the formal solution of Eq.~\eqref{Langevin} given by 
\begin{multline}\label{FormalSol}
x(t)=e^{-\mu\kappa t}x_0+v\int_{0}^{t}ds\, e^{-\mu\kappa(t-s)}\sigma(s)+\\
\int_{0}^{t}ds\, e^{-\mu\kappa(t-s)}\xi(s),
\end{multline}
with $x_0$ denotes the initial position.
The Fourier representation of the Dirac's delta function allows to write the $P(x,t)$ as
\begin{equation}
P(x,t)=\frac{1}{2\pi}\int_{-\infty}^\infty\Bigl\llangle e^{-ik(x-x(t))}\Bigr\rrangle_{\sigma,\xi},
\end{equation}
which after substitution of \eqref{FormalSol} and some rearrangements we get
\begin{multline}
P(x,t)=\frac{1}{2\pi}\int_{-\infty}^\infty e^{-ikx}\biggl[\exp\bigl\{ike^{-\mu\kappa t}x_0\bigr\}\times\\
\biggl\langle\exp\biggl\{ik\int_0^t ds e^{-\mu\kappa(t-s)}\xi(s)\biggr\}\biggr\rangle_\xi\biggr]\times
\\\biggl\langle\exp\biggl\{ik\int_0^tdse^{-\mu\kappa(t-s}\sigma(s)\biggr\}\biggr\rangle_\sigma,
\end{multline}
which corresponds to the convolution of the inverse Fourier transform of the Ornstein-Uhlenbeck propagator that appears in \eqref{TotalP}
\begin{subequations}
\begin{multline}
\hat{G}_\text{OU}(k,t\vert x_0)=\exp\{ike^{-\mu\kappa t}x_0\}\times\\
\biggl\langle\exp\biggl\{ik\int_0^t ds e^{-\mu\kappa(t-s)}\xi(s)\biggr\}\biggr\rangle_\xi,
\end{multline}
and of the inverse Fourier transform of the probability density of confined active motion
\begin{equation}
\hat{p}(k,t)=\biggl\langle\exp\biggl\{ik\int_0^tdse^{-\mu\kappa(t-s}\sigma(s)\biggr\}\biggr\rangle_\sigma,
\end{equation}
\end{subequations}
given in Eq.~\eqref{TotalP}. This completes the proof. 

\section{The moments of $P_\text{st}(x)$}
\label{sect:moments}
The characteristic function $\langle e^{-ikx}\rangle_{P_\text{st}}$, also know as the moments generating function $\hat{P}_\text{st}(k)=\int_{-\infty}^{\infty}dx\, e^{-ikx}P_\text{st}(x)$ has been computed explicitly as
\begin{equation}\label{CaracteristicFunction}
\hat{P}_\text{st}(k)=\hat{G}_\text{st}(k)\hat{p}_\text{st}(k),
\end{equation} 
with
\begin{equation*}
\hat{G}_\text{st}(k)=\exp\Bigl\{-\frac{D_T}{2\mu\kappa}k^2\Bigr\}
\end{equation*}
and
\begin{align*}
\hat{p}_\text{st}(k)=
&\frac{\Gamma\bigl(\ell+\frac{1}{2}\bigr)}{\sqrt{\pi}l_\text{trap}\Gamma(\ell)}\int_{-l_\text{trap}} ^{l_\text{trap}} \hspace{-0.5cm}dx\cos(kx)\biggl(1-\frac{x^{2}}{l_\text{trap}^{2}}\biggr)^{\ell-1}\\
=& _0F_1\Bigl(;\ell+\frac{1}{2};-\frac{k^2 l_\text{trap}^{2}}{4}\Bigr).
\end{align*}

The moments of $P_\text{st}(x)$ are given by the formula
\begin{equation}\label{Moments}
\bigl\langle x^{n}\bigr\rangle_{P_\text{st}}=(-i)^{n}\biggl[\frac{d^{n}}{dk^{n}}\hat{P}_\text{st}(k)\biggr]_{k=0}
\end{equation}
which after use of \eqref{CaracteristicFunction} and of the Leibniz formula for the $n$-th derivative of the product of two functions we get
\begin{equation}
\label{GralMoments}
\bigl\langle x^{n}\bigr\rangle_{P_\text{st}}=(-i)^{n}\sum_{m=0}^{n}\binom{n}{m}\biggl[\frac{d^{m}}{dk^{m}}\hat{G}_\text{st}(k)
\frac{d^{n-m}}{dk^{n-m}}\hat{p}_\text{st}(k)\biggr]_{k=0}.
\end{equation} 
Odd moments vanishes due to the even symmetry with respect to the origin. 

Being a Gaussian, $\hat{G}_\text{st}(k)$, leads to the well-known formulas for the even moments of $G_\text{st}(x)$
\begin{equation}
\bigl\langle x^{2n}\bigr\rangle_{G_\text{st}}=(2n-1)!!\biggl(\frac{D_T}{\mu\kappa}\biggr)^{n},
\end{equation}
while the even moments of $p_\text{st}(x)$ are given by
\begin{equation}
\bigl\langle x^{2n}\bigr\rangle_{p_\text{st}}=\frac{\Gamma\bigl(\ell+\frac{1}{2}\bigr)\Gamma\bigl(n+\frac{1}{2}\bigr)}{\Gamma\bigl(\ell+n+\frac{1}{2}\bigr)\Gamma\bigl(n+1\bigr)}\frac{l_\text{trap}^{2n}}{2\sqrt{\pi}}.
\end{equation}

\subsection{The second moment}
From Eq.~\eqref{GralMoments}, the second moment of $P_\text{st}$ is given by
\begin{subequations}
\begin{equation}
\langle x^{2}\rangle_{P_\text{st}}=\langle x^{2}\rangle_{G_\text{st}}+\langle x^{2}\rangle_{p_\text{st}},
\end{equation}
with 
\begin{subequations}
\begin{align}
&\langle x^{2}\rangle_{G_\text{st}}=\frac{D_T}{\mu\kappa},\\
&\langle x^{2}\rangle_{p_\text{st}}=\frac{1}{\bigl(\ell+\frac{1}{2}\bigr)}\biggl(\frac{l_\text{trap}}{2}\biggr)^2.
\end{align}
\end{subequations}

\subsection{The fourth moment of $P_\text{st}$}
From Eq.~\eqref{GralMoments}, the fourth moment is
\begin{equation}
\langle x^{4}\rangle_{P_\text{st}}=\langle x^{4}\rangle_{G_\text{st}}+6\langle x^{2}\rangle_{G_\text{st}}\langle x^{2}\rangle_{p_\text{st}}
+\langle x^{4}\rangle_{p_\text{st}},
\end{equation}
\end{subequations}
with
\begin{subequations}
\begin{align}
&\langle x^{2}\rangle_{G_\text{st}}=\frac{D_T}{\mu\kappa},\\
&\langle x^{4}\rangle_{G_\text{st}}=3\biggl(\frac{D_T}{\mu\kappa}\biggr)^{2},\\
&\langle x^{2}\rangle_{p_\text{st}}=\frac{1}{\bigl(\ell+\frac{1}{2}\bigr)}\biggl(\frac{l_\text{trap}}{2}\biggr)^2,\\
&\langle x^{4}\rangle_{p_\text{st}}=\frac{3}{\bigl(\ell+\frac{1}{2}\bigr)\bigl(\ell+\frac{3}{2}\bigr)}\biggl(\frac{l_\text{trap}}{2}\biggr)^4.
\end{align}
\end{subequations}

\begin{acknowledgments}
This work was supported by UNAM-PAPIIT IN110626.
FJS kindly thanks Hugo Uriel Granados Torres for his interest in the initial part of this work. 
\end{acknowledgments}

\section*{Data Availability Statements}
The raw data supporting the conclusions of this article will be made available by the authors on request.


\begin{thebibliography}{22}
\expandafter\ifx\csname natexlab\endcsname\relax\def\natexlab#1{#1}\fi
\expandafter\ifx\csname bibnamefont\endcsname\relax
  \def\bibnamefont#1{#1}\fi
\expandafter\ifx\csname bibfnamefont\endcsname\relax
  \def\bibfnamefont#1{#1}\fi
\expandafter\ifx\csname citenamefont\endcsname\relax
  \def\citenamefont#1{#1}\fi
\expandafter\ifx\csname url\endcsname\relax
  \def\url#1{\texttt{#1}}\fi
\expandafter\ifx\csname urlprefix\endcsname\relax\def\urlprefix{URL }\fi
\providecommand{\bibinfo}[2]{#2}
\providecommand{\eprint}[2][]{\url{#2}}

\bibitem[{\citenamefont{Schnitzer}(1993)}]{SchnitzerPhysRevRE1993}
\bibinfo{author}{\bibfnamefont{M.~J.} \bibnamefont{Schnitzer}},
  \bibinfo{journal}{Phys. Rev. E} \textbf{\bibinfo{volume}{48}},
  \bibinfo{pages}{2553} (\bibinfo{year}{1993}),
  \urlprefix\url{http://link.aps.org/doi/10.1103/PhysRevE.48.2553}.

\bibitem[{\citenamefont{Sevilla et~al.}(2019)\citenamefont{Sevilla, Arzola, and
  Cital}}]{SevillaPRE2019}
\bibinfo{author}{\bibfnamefont{F.~J.} \bibnamefont{Sevilla}},
  \bibinfo{author}{\bibfnamefont{A.~V.} \bibnamefont{Arzola}},
  \bibnamefont{and} \bibinfo{author}{\bibfnamefont{E.~P.} \bibnamefont{Cital}},
  \bibinfo{journal}{Phys. Rev. E} \textbf{\bibinfo{volume}{99}},
  \bibinfo{pages}{012145} (\bibinfo{year}{2019}),
  \urlprefix\url{https://link.aps.org/doi/10.1103/PhysRevE.99.012145}.

\bibitem[{\citenamefont{Dhar et~al.}(2019)\citenamefont{Dhar, Kundu, Majumdar,
  Sabhapandit, and Schehr}}]{DharPhysRevE2019}
\bibinfo{author}{\bibfnamefont{A.}~\bibnamefont{Dhar}},
  \bibinfo{author}{\bibfnamefont{A.}~\bibnamefont{Kundu}},
  \bibinfo{author}{\bibfnamefont{S.~N.} \bibnamefont{Majumdar}},
  \bibinfo{author}{\bibfnamefont{S.}~\bibnamefont{Sabhapandit}},
  \bibnamefont{and} \bibinfo{author}{\bibfnamefont{G.}~\bibnamefont{Schehr}},
  \bibinfo{journal}{Phys. Rev. E} \textbf{\bibinfo{volume}{99}},
  \bibinfo{pages}{032132} (\bibinfo{year}{2019}),
  \urlprefix\url{https://link.aps.org/doi/10.1103/PhysRevE.99.032132}.

\bibitem[{\citenamefont{Garcia-Millan and
  Pruessner}(2021)}]{Garcia-MillanJStatMech2021}
\bibinfo{author}{\bibfnamefont{R.}~\bibnamefont{Garcia-Millan}}
  \bibnamefont{and}
  \bibinfo{author}{\bibfnamefont{G.}~\bibnamefont{Pruessner}},
  \bibinfo{journal}{Journal of Statistical Mechanics: Theory and Experiment}
  \textbf{\bibinfo{volume}{2021}}, \bibinfo{pages}{063203}
  (\bibinfo{year}{2021}),
  \urlprefix\url{https://dx.doi.org/10.1088/1742-5468/ac014d}.

\bibitem[{\citenamefont{Crisanti and Paoluzzi}(2023)}]{CrisantiPhysRevE2023}
\bibinfo{author}{\bibfnamefont{A.}~\bibnamefont{Crisanti}} \bibnamefont{and}
  \bibinfo{author}{\bibfnamefont{M.}~\bibnamefont{Paoluzzi}},
  \bibinfo{journal}{Phys. Rev. E} \textbf{\bibinfo{volume}{107}},
  \bibinfo{pages}{034110} (\bibinfo{year}{2023}),
  \urlprefix\url{https://link.aps.org/doi/10.1103/PhysRevE.107.034110}.

\bibitem[{\citenamefont{Meyer and Metzler}(2023)}]{MeyerNewJPhys2023}
\bibinfo{author}{\bibfnamefont{P.~G.} \bibnamefont{Meyer}} \bibnamefont{and}
  \bibinfo{author}{\bibfnamefont{R.}~\bibnamefont{Metzler}},
  \bibinfo{journal}{New Journal of Physics} \textbf{\bibinfo{volume}{25}},
  \bibinfo{pages}{063003} (\bibinfo{year}{2023}),
  \urlprefix\url{https://dx.doi.org/10.1088/1367-2630/acd94f}.

\bibitem[{\citenamefont{Frydel}(2023)}]{FrydelPoF2023}
\bibinfo{author}{\bibfnamefont{D.}~\bibnamefont{Frydel}},
  \bibinfo{journal}{Physics of Fluids} \textbf{\bibinfo{volume}{35}},
  \bibinfo{pages}{101905} (\bibinfo{year}{2023}), ISSN
  \bibinfo{issn}{1070-6631},
  \eprint{https://pubs.aip.org/aip/pof/article-pdf/doi/10.1063/5.0173374/18165520/101905\_1\_5.0173374.pdf},
  \urlprefix\url{https://doi.org/10.1063/5.0173374}.

\bibitem[{\citenamefont{Dutta et~al.}(2024)\citenamefont{Dutta, Kundu,
  Sabhapandit, and Basu}}]{DuttaPhysRevE2024}
\bibinfo{author}{\bibfnamefont{D.}~\bibnamefont{Dutta}},
  \bibinfo{author}{\bibfnamefont{A.}~\bibnamefont{Kundu}},
  \bibinfo{author}{\bibfnamefont{S.}~\bibnamefont{Sabhapandit}},
  \bibnamefont{and} \bibinfo{author}{\bibfnamefont{U.}~\bibnamefont{Basu}},
  \bibinfo{journal}{Phys. Rev. E} \textbf{\bibinfo{volume}{110}},
  \bibinfo{pages}{044107} (\bibinfo{year}{2024}),
  \urlprefix\url{https://link.aps.org/doi/10.1103/PhysRevE.110.044107}.
  
  \bibitem[{\citenamefont{Gu\'eneau et~al.}(2026)\citenamefont{Gu\'eneau,
  Majumdar, and Schehr}}]{GueneauPhysRevE2026}
\bibinfo{author}{\bibfnamefont{M.}~\bibnamefont{Gu\'eneau}},
  \bibinfo{author}{\bibfnamefont{S.~N.} \bibnamefont{Majumdar}},
  \bibnamefont{and} \bibinfo{author}{\bibfnamefont{G.}~\bibnamefont{Schehr}},
  \bibinfo{journal}{Phys. Rev. E} \textbf{\bibinfo{volume}{114}},
  \bibinfo{pages}{014144} (\bibinfo{year}{2026}),
  \urlprefix\url{https://link.aps.org/doi/10.1103/mlkm-vgbd}.

\bibitem[{\citenamefont{Smith}(2023)}]{SmithPhysRevE2023}
\bibinfo{author}{\bibfnamefont{N.~R.} \bibnamefont{Smith}},
  \bibinfo{journal}{Phys. Rev. E} \textbf{\bibinfo{volume}{108}},
  \bibinfo{pages}{L022602} (\bibinfo{year}{2023}),
  \urlprefix\url{https://link.aps.org/doi/10.1103/PhysRevE.108.L022602}.

\bibitem[{\citenamefont{Farago and Smith}(2024)}]{FaragoPhysRevE2024}
\bibinfo{author}{\bibfnamefont{O.}~\bibnamefont{Farago}} \bibnamefont{and}
  \bibinfo{author}{\bibfnamefont{N.~R.} \bibnamefont{Smith}},
  \bibinfo{journal}{Phys. Rev. E} \textbf{\bibinfo{volume}{109}},
  \bibinfo{pages}{044121} (\bibinfo{year}{2024}),
  \urlprefix\url{https://link.aps.org/doi/10.1103/PhysRevE.109.044121}.

\bibitem[{\citenamefont{Singh and Farago}(2025)}]{SinghPhysRevE2025}
\bibinfo{author}{\bibfnamefont{R.~K.} \bibnamefont{Singh}} \bibnamefont{and}
  \bibinfo{author}{\bibfnamefont{O.}~\bibnamefont{Farago}},
  \bibinfo{journal}{Phys. Rev. E} \textbf{\bibinfo{volume}{111}},
  \bibinfo{pages}{064131} (\bibinfo{year}{2025}),
  \urlprefix\url{https://link.aps.org/doi/10.1103/c51r-fmgw}.

\bibitem[{\citenamefont{{Takatori Sho C.} et~al.}(2016)\citenamefont{{Takatori
  Sho C.}, {De Dier Raf}, {Vermant Jan}, and {Brady John
  F.}}}]{TakatoriNatcomms2016}
\bibinfo{author}{\bibnamefont{{Takatori Sho C.}}},
  \bibinfo{author}{\bibnamefont{{De Dier Raf}}},
  \bibinfo{author}{\bibnamefont{{Vermant Jan}}}, \bibnamefont{and}
  \bibinfo{author}{\bibnamefont{{Brady John F.}}}, \bibinfo{journal}{Nature
  Communications} \textbf{\bibinfo{volume}{7}}, \bibinfo{pages}{10694}
  (\bibinfo{year}{2016}),
  \urlprefix\url{http://www.nature.com/articles/ncomms10694\#supplementary-information}.

\bibitem[{\citenamefont{Argun et~al.}(2016)\citenamefont{Argun, Moradi,
  Pin\c{c}e, Bagci, Imparato, and Volpe}}]{ArgunPRE2016}
\bibinfo{author}{\bibfnamefont{A.}~\bibnamefont{Argun}},
  \bibinfo{author}{\bibfnamefont{A.-R.} \bibnamefont{Moradi}},
  \bibinfo{author}{\bibfnamefont{E.}~\bibnamefont{Pin\c{c}e}},
  \bibinfo{author}{\bibfnamefont{G.~B.} \bibnamefont{Bagci}},
  \bibinfo{author}{\bibfnamefont{A.}~\bibnamefont{Imparato}}, \bibnamefont{and}
  \bibinfo{author}{\bibfnamefont{G.}~\bibnamefont{Volpe}},
  \bibinfo{journal}{Phys. Rev. E} \textbf{\bibinfo{volume}{94}},
  \bibinfo{pages}{062150} (\bibinfo{year}{2016}),
  \urlprefix\url{https://link.aps.org/doi/10.1103/PhysRevE.94.062150}.

\bibitem[{\citenamefont{Wexler et~al.}(2020)\citenamefont{Wexler, Gov,
  Rasmussen, and Bel}}]{WexlerPhysRevResearch2020}
\bibinfo{author}{\bibfnamefont{D.}~\bibnamefont{Wexler}},
  \bibinfo{author}{\bibfnamefont{N.}~\bibnamefont{Gov}},
  \bibinfo{author}{\bibfnamefont{K.~O.} \bibnamefont{Rasmussen}},
  \bibnamefont{and} \bibinfo{author}{\bibfnamefont{G.}~\bibnamefont{Bel}},
  \bibinfo{journal}{Phys. Rev. Res.} \textbf{\bibinfo{volume}{2}},
  \bibinfo{pages}{013003} (\bibinfo{year}{2020}),
  \urlprefix\url{https://link.aps.org/doi/10.1103/PhysRevResearch.2.013003}.

\bibitem[{\citenamefont{Schmidt et~al.}(2021)\citenamefont{Schmidt,
  {\v{S}}{\'\i}pov{\'a}-Jungov{\'a}, K{\"a}ll, W{\"u}rger, and
  Volpe}}]{SchmidtNatureComm2021}
\bibinfo{author}{\bibfnamefont{F.}~\bibnamefont{Schmidt}},
  \bibinfo{author}{\bibfnamefont{H.}~\bibnamefont{{\v{S}}{\'\i}pov{\'a}-Jungov{\'a}}},
  \bibinfo{author}{\bibfnamefont{M.}~\bibnamefont{K{\"a}ll}},
  \bibinfo{author}{\bibfnamefont{A.}~\bibnamefont{W{\"u}rger}},
  \bibnamefont{and} \bibinfo{author}{\bibfnamefont{G.}~\bibnamefont{Volpe}},
  \bibinfo{journal}{Nature Communications} \textbf{\bibinfo{volume}{12}},
  \bibinfo{pages}{1902} (\bibinfo{year}{2021}),
  \urlprefix\url{https://doi.org/10.1038/s41467-021-22187-z}.

\bibitem[{\citenamefont{Buttinoni et~al.}(2022)\citenamefont{Buttinoni,
  Caprini, Alvarez, Schwarzendahl, and Löwen}}]{ButtinoniEPL2022}
\bibinfo{author}{\bibfnamefont{I.}~\bibnamefont{Buttinoni}},
  \bibinfo{author}{\bibfnamefont{L.}~\bibnamefont{Caprini}},
  \bibinfo{author}{\bibfnamefont{L.}~\bibnamefont{Alvarez}},
  \bibinfo{author}{\bibfnamefont{F.~J.} \bibnamefont{Schwarzendahl}},
  \bibnamefont{and} \bibinfo{author}{\bibfnamefont{H.}~\bibnamefont{Löwen}},
  \bibinfo{journal}{Europhysics Letters} \textbf{\bibinfo{volume}{140}},
  \bibinfo{pages}{27001} (\bibinfo{year}{2022}),
  \urlprefix\url{https://dx.doi.org/10.1209/0295-5075/ac9c28}.

\bibitem[{\citenamefont{Caraglio and Franosch}(2022)}]{CaraglioPhysRevLett2022}
\bibinfo{author}{\bibfnamefont{M.}~\bibnamefont{Caraglio}} \bibnamefont{and}
  \bibinfo{author}{\bibfnamefont{T.}~\bibnamefont{Franosch}},
  \bibinfo{journal}{Phys. Rev. Lett.} \textbf{\bibinfo{volume}{129}},
  \bibinfo{pages}{158001} (\bibinfo{year}{2022}),
  \urlprefix\url{https://link.aps.org/doi/10.1103/PhysRevLett.129.158001}.

\bibitem[{\citenamefont{Smith et~al.}(2022)\citenamefont{Smith, Le~Doussal,
  Majumdar, and Schehr}}]{SmithPhysRevE2022}
\bibinfo{author}{\bibfnamefont{N.~R.} \bibnamefont{Smith}},
  \bibinfo{author}{\bibfnamefont{P.}~\bibnamefont{Le~Doussal}},
  \bibinfo{author}{\bibfnamefont{S.~N.} \bibnamefont{Majumdar}},
  \bibnamefont{and} \bibinfo{author}{\bibfnamefont{G.}~\bibnamefont{Schehr}},
  \bibinfo{journal}{Phys. Rev. E} \textbf{\bibinfo{volume}{106}},
  \bibinfo{pages}{054133} (\bibinfo{year}{2022}),
  \urlprefix\url{https://link.aps.org/doi/10.1103/PhysRevE.106.054133}.

\bibitem[{\citenamefont{Szamel}(2023)}]{SzamelPhysRevE2023}
\bibinfo{author}{\bibfnamefont{G.}~\bibnamefont{Szamel}},
  \bibinfo{journal}{Phys. Rev. E} \textbf{\bibinfo{volume}{107}},
  \bibinfo{pages}{054602} (\bibinfo{year}{2023}),
  \urlprefix\url{https://link.aps.org/doi/10.1103/PhysRevE.107.054602}.

\bibitem[{\citenamefont{Baldovin et~al.}(2023)\citenamefont{Baldovin,
  Gu\'ery-Odelin, and Trizac}}]{BaldovinPhysRevLett2023}
\bibinfo{author}{\bibfnamefont{M.}~\bibnamefont{Baldovin}},
  \bibinfo{author}{\bibfnamefont{D.}~\bibnamefont{Gu\'ery-Odelin}},
  \bibnamefont{and} \bibinfo{author}{\bibfnamefont{E.}~\bibnamefont{Trizac}},
  \bibinfo{journal}{Phys. Rev. Lett.} \textbf{\bibinfo{volume}{131}},
  \bibinfo{pages}{118302} (\bibinfo{year}{2023}),
  \urlprefix\url{https://link.aps.org/doi/10.1103/PhysRevLett.131.118302}.

\bibitem[{\citenamefont{Frydel}(2024)}]{FrydelPoF2024}
\bibinfo{author}{\bibfnamefont{D.}~\bibnamefont{Frydel}},
  \bibinfo{journal}{Physics of Fluids} \textbf{\bibinfo{volume}{36}},
  \bibinfo{pages}{011910} (\bibinfo{year}{2024}), ISSN
  \bibinfo{issn}{1070-6631},
  \eprint{https://pubs.aip.org/aip/pof/article-pdf/doi/10.1063/5.0179375/18573177/011910\_1\_5.0179375.pdf},
  \urlprefix\url{https://doi.org/10.1063/5.0179375}.

\bibitem[{\citenamefont{Sun et~al.}(2025)\citenamefont{Sun, Ye, and
  Podgornik}}]{SunPhysRevE2025}
\bibinfo{author}{\bibfnamefont{A.}~\bibnamefont{Sun}},
  \bibinfo{author}{\bibfnamefont{F.}~\bibnamefont{Ye}}, \bibnamefont{and}
  \bibinfo{author}{\bibfnamefont{R.}~\bibnamefont{Podgornik}},
  \bibinfo{journal}{Phys. Rev. E} \textbf{\bibinfo{volume}{111}},
  \bibinfo{pages}{044136} (\bibinfo{year}{2025}),
  \urlprefix\url{https://link.aps.org/doi/10.1103/PhysRevE.111.044136}.

\end{thebibliography}
\end{document}